\documentclass[prd,twocolumn,showpacs,showkeys,preprintnumbers,floatfix,nofootinbib]{revtex4-1}
\usepackage{amsmath, amssymb, amsthm, color,latexsym,graphicx,array,multirow}
\usepackage{dcolumn} 
\usepackage[caption=false, labelformat=empty]{subfig} %

\newcommand{\gev}{\text{GeV}}

\newcommand{\kt}{k_T}
\newcommand{\CA}{\text{C/A}}
\newcommand{\Akt}{\text{anti-}k_{T} }
\newcommand{\pjsp}[1]{$\text{PJSP}\:\mathbf{#1}$}

\begin{document}

\title{Phenomenology of Photon-Jets}
\author{ Stephen D. Ellis }
\author{ Tuhin S. Roy }
\author{ Jakub Scholtz  }
\affiliation{Physics Department, University of Washington,
 Seattle, WA 98195-1560, USA \\
	}

\date{\today}
\begin{abstract}

One of the challenges of collider physics is to unambiguously associate detector based objects with the corresponding elementary physics objects.	 A particular example is the association of calorimeter-based objects such as ``jets'', identified with a standard (IR-safe) jet algorithm, with the underlying physics objects, which may be QCD-jets (arising from a scattered parton), electrons, photons and, as discussed here, photon-jets (a group of collinear photons).  This separation is especially interesting in the context of Higgs searches, where the signal includes both di-photon (in the Standard Model) and di-photon-jet decays (in a variety of Beyond the Standard Model scenarios), while QCD provides an ever-present background.  Here we describe the implementation of techniques from the rapidly evolving area of jet-substructure studies to not only enhance the more familiar photon-QCD separation, but also separately distinguish photon-jets, i.e., separate usual jets into three categories: single photons, photon-jets and QCD.  The efficacy of these techniques for separation is illustrated through studies of simulated data.  
	 
\end{abstract}

\maketitle

\section{Introduction}

The  Large Hadron Collider (LHC) has clearly exhibited its ability to make discoveries with the observation of a new resonance~\cite{:2012gk,:2012gu} with even spin that decays to photons and Z bosons as expected of the Standard Model (SM) Higgs particle.  Thus precise measurements of the decays of this resonance into various channels (whether standard or not),  are of the utmost importance.  At the same time, it is essential to verify our understanding of the existing channels, in particular,  $h \rightarrow \gamma \gamma$.  How well are these photons defined?  Can physics objects other than single photons leave signatures in the detector similar to that of a photon?  Not surprisingly, the answer is yes~\cite{Dobrescu:2000jt, Toro:2012sv, Draper:2012xt}.  Given the granularity of the calorimeters, an object consisting of (nearly) collinear photons, typically labeled a photon-jet, will generate a  signature  similar to that of a single photon.
The possibility that the Higgs particle decays to multiple collinear photons is not new~\cite{Dobrescu:2000jt, Draper:2012xt}.  Simple models where the Higgs decays to almost massless scalars that each in turn decay to a pair of photons, typically do not give rise to events with four separately identifiable photons, but rather to pairs of photon-jets, each with 2 photons.   Slightly more complicated models can produce Higgs decays to photon-jets with $4, 6, \cdots$ photons. We will discuss concrete models where the Higgs decays to  photon-jets with $2$ and $4$ photons per photon-jet.  Thus it is essential to develop tools to separate single photons from photon-jets from QCD-jets.  Otherwise we are unlikely to understand either the signal or the background.

ATLAS recently made attempts to identify photon-jets from Higgs decays~\cite{ATLAS-CONF-2012-079}. These analyses rely on relaxing the isolation/shower shape criteria, which use the differing distributions of energy deposition within the calorimeter cells to quite successfully discriminate single photons from QCD-jets.  Unfortunately, the parameters of the underlying model can be easily adjusted so that the resultant photon-jets pass the strictest isolation/shower shape criteria just like photons. More importantly, loosening isolation criteria results in a larger fake rate for QCD-jets. Discriminating photon-jets from QCD-jets is more challenging than separating single photons from QCD-jets.
 
Fortunately jet substructure techniques~\cite{Seymour:1993mx, Brooijmans:1077731,Butterworth:2007ke, Butterworth:2008iy, Thaler:2008ju, Kaplan:2008ie} have recently been developed to distinguish QCD-jets from jets containing boosted heavy particle decays, and we can use this work for detection of photon-jets.  More broadly, `jets', as defined by an infrared safe jet clustering algorithm, are being proposed as a universal language to describe \textit{all} calorimeter objects including single photons, photon-jets and QCD-jets.  By using the tools developed in jet substructure physics, we do not need to rely on isolation cuts. We supplement the traditional/conventional variables currently used to discriminate photons from QCD-jets with substructure variables that probe in detail the energy distribution within the jet. Note that the photons-jets are composed of energetic photons distributed inside the jet, where the distribution is a result of the kinematic features of the model, e.g., the masses and spins of the intermediate particles. The existence of this structure within photon-jet suggests that substructure variables will be efficient at finding and discriminating photon-jets.  We show that our analysis is capable of separating  photon-jets from both single photons and QCD-jets  \emph{at least as} efficiently as the traditional discriminators separate photons from QCD-jets. 

There is another important advantage to applying jet substructure techniques to purely electro-magnetic calorimeter (ECal) objects.  The introduction of `grooming' algorithms (including filtering~\cite{Butterworth:2008iy, Butterworth:2008sd, Butterworth:2008tr}, pruning~\cite{Ellis:2009su, Ellis:2009me}, and trimming~\cite{Krohn:2009th}) promised to suppress the undesirable contributions to purely hadronic jets from the underlying event (the largely uncorrelated soft interactions surrounding the interesting hard scattering) and from pile-up (the truly uncorrelated proton-proton collisions that occur in the same time window).  Indeed, the recent results from studies at  ATLAS~\cite{ATLAS-CONF-2012-065} and CMS~\cite{CMS-PAS-EXO-11-095} indicate this grooming is effective. We expect that this substructure-based grooming will work as well for all ECal based objects.

It should be noted that in the context of Higgs physics, the decay to photon-jets is not the only example where the collinearity of the decay products adds complexity to the analysis. Collinearity plays a role for traditional decays of the Higgs boson when it is boosted. In Ref.~\cite{Butterworth:2008iy}, the authors exploited the collinearity of the $b$-quarks in boosted Higgs decays (both quarks in a single jet) to greatly enhance the chances of detecting the $h\rightarrow b\bar{b}$ channel, featuring jet substructure as a mainstream tool (see also Refs.~\cite{Seymour:1993mx, Brooijmans:1077731,Butterworth:2007ke}).   The application of jet substructure in Higgs physics has now become a very active area of research, applied both to the SM Higgs~\cite{Plehn:2009rk,Gallicchio:2010dq, Hackstein:2010wk} as well to beyond the SM Higgs scenarios~\cite{Kribs:2009yh, Kribs:2010hp, Kribs:2010ii,Katz:2010iq, Englert:2011iz,Son:2012mb}.  For reviews, more detailed descriptions, and references see Refs.~\cite{Abdesselam:2010pt, Altheimer:2012mn}.

The paper is organized as follows: in Sec.~\ref{sec:simplified_models}, we start with a simplified model for photon-jets. We propose a set of benchmark points, where we take different combinations of masses and parameters in the simplified model to produce photon-jets displaying a variety of distinct kinematics. In Sec.~\ref{sec:simulations} we define the details of our simulation. We describe, in detail, how we generate  samples of photon-jets, one for each of the benchmark points, QCD-jets, and single photons.  We present our analysis in Sec.~\ref{sec:analysis}. We describe all the variables that  we use in this work to discriminate photon-jets from QCD-jets from single photons. Then we combine these variables in a multivariate analysis. We train boosted decision trees (BDTs) using the samples of jets and use these to optimize the discriminating power of our analyses. We also show how these BDTs can be used to simultaneously separate photon-jets, photons, and QCD-jets from each other.  Our conclusions are presented in Sec.~\ref{sec:conclusion}. 

\section{\label{sec:simplified_models} Simple Model for Photon-Jets}

By definition, photons-jets refer to calorimeter objects consisting of more than one hard photon. However, such a broad definition presents a challenge since all  photon-jets are not the same. They differ in terms of the number of hard constituent photons as well as in the distribution of those photons within the photon-jet.  To provide a systematic phenomenological study of photon-jets we classify these objects in more detail in terms of the production mechanism and consider a broad range.  We will refer to the various production scenarios as `benchmark' scenarios.  
We find that a simple model in the spirit of Ref.~\cite{Alves:2011wf} with two new particles is sufficient to characterize these benchmarks.   The model includes a small number of interactions and we can vary the strength of these interaction and the new particle masses in order to generate the benchmark scenarios.  In particular,  we introduce two scalar fields $n_1$ and $n_2$ of mass $m_1$ and $m_2$ respectively.  Without loss of generality, we choose the naming convention such that $m_1 > m_2$.  Neither $n_1$ nor $n_2$ carry any SM charges. We use the following interactions to generate photon-jets
\begin{equation}
	\label{eq:simplified-model}
	 \frac{1}{2} \mu_h  \: h n_1^2 +  \frac{1}{2} \mu_{12} \: n_1 n_2^2  + \left( \frac{\eta_1}{m_1} \: n_1 
	      + \frac{\eta_2}{m_2}  \: n_2\right) F^{\mu\nu}F_{\mu\nu}  \; , 
\end{equation}
where $\mu_h, \mu_{12}$ are mass parameters,  $\eta_1, \eta_2$ are dimensionless coupling constants, and $F_{\mu\nu}$ is the electromagnetic field strength operator.
 
This simple model bears a resemblance to a Higgs portal scenario~\cite{Schabinger:2005ei,Patt:2006fw,Strassler:2006im} because of the $\mu_h$ coupling. In the Higgs portal language, $n_1$ and $n_2$ constitute a `hidden' sector while the coupling $\mu_h$ provides a tunnel to the corresponding `hidden valley'. The electromagnetic couplings (proportional to the $\eta$ parameters) provide ways for the new particles to decay back to SM particles, photons in this case.  With respect to Higgs physics, this simple model provides a realistic example where the SM Higgs field decays through the new particles to multiple photons.  In the limit $m_1 \ll m_h$, the resultant photons (the decay products of $n_1$) are essentially collinear. 

In Table~\ref{table:bench} we list the benchmark scenarios (labeled photon-jet study points or PJSPs) that we investigate in this work.  All are generated by varying the parameters in Eq.\eqref{eq:simplified-model}.  The symbol $\text{X}$ in Table~\ref{table:bench} denotes that a non-zero value is selected for that parameter, which then determines the decay mode. We have chosen the benchmarks in such a way that the parameters denoted by $\text{X}$ only change the \textit{total} width of the decaying particles. As long as the decays are prompt, the exact values of these parameters are irrelevant to the phenomenological properties of the photon-jets.
\begin{table}[h]
	\centering
	\begin{tabular}{|c|c|c|c|c|c|}
		\hline
		\multirow{2}{*}{Study Points} & $m_1$  & $m_2$  & $\mu_{12}$  
							& \multirow{2}{*}{$\eta_1$}& \multirow{2}{*}{$\eta_2$} \\
			    & $\left( \gev \right)$ & $\left( \gev \right)$ &  $\left( \gev \right)$ &  & \\
		\hline   
		\pjsp{1} & $0.5$ 	& 		& \multirow{3}{*}{$0$}	&  \multirow{3}{*}{$ \text{X}$} & \\
		\pjsp{2} & $1.0$  	& 		& 	&	& \\
		\pjsp{3} & $10.0$	& 		& 	&	& \\
		\hline
		\pjsp{4} & $2.0$	& $0.5$	& \multirow{5}{*}{$\text{X}$}	
						& \multirow{5}{*}{$0$}	& \multirow{5}{*}{$\text{X}$} \\
		\cline{1-3}
		\pjsp{5} &  \multirow{2}{*}{$5.0$}				& $0.5$	& 	&	&\\   
		\pjsp{6} &  				& $1.0$	& 	&	& \\   
		\cline{1-3}
		\pjsp{7} &  \multirow{2}{*}{$10.0$}			& $0.5$&   	&	& \\   
		\pjsp{8} &  				& $1.0$		& 	&	& \\   
		\hline
	\end{tabular}
	\caption{\label{table:bench} The study points used in our analysis. For \pjsp{1-3}, $n_2$ does not participate in the decay chain since $\mu_{12} = 0$ and the $m_2$ and $\eta_2$ columns are empty. By $\text{X}$ we denote that a non-zero value is chosen for the parameter, which facilitates prompt decays, but the specific value plays no role.  }
\end{table}
In all these study points we take the Higgs particle to decay to a pair of $n_1$ particles. The small $n_1$ mass ($m_1 \ll m_h$) ensures that the decay products of the $n_1$ are highly collimated.  In the Higgs particle rest frame, which is close to the laboratory frame on average, each $n_1$ has momentum $\sim m_h/2$ and the typical angular separation between the $n_1$ decay products is of the order of $4 m_1/m_h $.  Note that, given we always consider $m_1 \leq 10~\gev$, we expect the typical angular separation between the $n_1$ decay products to be $\lesssim 1/3$ (we use $m_h = 120~\gev$). As long as the angular size of photon-jets is larger than $1/3$, we expect to capture all the decay products of the $n_1$ in each photon-jet for all the  benchmark points.  

For the study points \pjsp{1-3} the mass parameter $\mu_{12}$ is set to zero and $n_1 \rightarrow \gamma \gamma$ is the only possible $n_1$ decay mode.  Hence these scenarios are characterized by photons-jets with typically $2$ hard photons per jet, and $n_2$ plays no role in the phenomenology (so no $n_2$ mass or coupling values are included in the table).   In these scenarios the Higgs particle cascade decays to four photons ($h \rightarrow n_1 n_1 \rightarrow \gamma \gamma\gamma \gamma$).  The precise value of $m_1$ governs the angular separation of the two photons inside the photon-jets.  For a very small $m_1$, each photon-jet looks much like a single photon. (Of course, if the Higgs is highly boosted, the decay results in a single photon-jet containing all 4 photons.)

For study points \pjsp{4-8} we set $\eta_1$ to zero and $\mu_{12}$ to a non-zero value.  In these contrasting scenarios the only $n_1$ decay mode involves the chain $n_1 \rightarrow n_2 n_2 \rightarrow \gamma \gamma \gamma \gamma $.  Hence the Higgs decays again to two photon-jets, but now each photon-jet typically contains four photons (the $n_1$ decay products).  (In this case, a highly boosted Higgs yields a single photon-jet containing 8 photons.)

\section{\label{sec:simulations} Simulation Details}

In order to generate samples of photons-jets, we implement the simple model of Eq.~\eqref{eq:simplified-model} in MadGraph~$5$~\cite{Alwall:2011uj}.  For each benchmark point we generate matrix elements corresponding to the process $pp \rightarrow h \rightarrow n_1n_1$ (via gluon fusion) using  MadGraph~$5$ with $m_h = 120~\gev$, which we employ as input to Pythia~$8.1$~\cite{Sjostrand:2006za,Sjostrand:2007gs} in order to generate the full events and for the subsequent $n_1$ decays.  Since the Higgs production is evaluated at lowest order, the produced Higgs particles have zero transverse momentum.  We use the QCD dijet events generated by standalone Pythia~$8.1$ to provide a sample of QCD-jets. In order to define a sample of single photons, we also generate $pp \rightarrow h \rightarrow \gamma \gamma$ events where the photons are well separated.   Finally, we include initial state radiation (ISR), final state radiation (FSR) and multiple parton interactions (MI, i.e., the UE) as implemented  in Pythia~$8.1$ to simulate the relevant busy hadronic collider environment. 

The Pythia output final states are subjected to our minimal detector simulation. In the following we describe briefly how we treat the final state particles in each event: 
\begin{itemize}

\item We identify all charged particles with transverse momentum $p_T > 2~\gev$ and pseudorapidity $|\eta| < 2.5$ as charged tracks.

\item In a real detector, tracks are also generated if photons convert within the pixel part of the tracker. In this work, we simulate this photon conversion process by associating with each photon a probability for it to convert in the tracker.~\footnote{ We do not simulate the magnetic field in the detector. Consequently the $e^+ e^-$ pairs from photon conversion continue in the direction of the photon. So for every converted photon we obtain effectively a single track, if the photon passes the $p_{T}$ threshold.}  

The probability is a function of the number of radiation lengths of material the photon has to traverse in order to escape the inner part of the tracker. We use the specifications of the ATLAS detector in order to model this pseudorapidity dependent probability distribution. The details of this procedure are outlined in the Appendix~\ref{sec:app-conversion}.

\item  In our simulation, all particles (except charged particles with $E < 0.1~\gev$) reach the calorimeters, and all of these (except muons with $p_T > 0.5~\gev$) deposit all of their energy in the calorimeters. The electromagnetic calorimeter (ECal) is modeled as cells of size $0.025\times0.025$ in ($\eta$-$\phi$), whereas the hadronic calorimeter (HCal) is taken to have more coarse granularity with $0.1\times0.1$ cells.  Besides photons and electrons, soft muons and soft hadrons (soft means $E < 0.5~\gev$) are treated as depositing all of their energy in the ECal.  More energetic hadrons are absorbed in the HCal, while more energetic muons escape the calorimeter. For a more detailed picture see Appendix~\ref{sec:Calorimeter}.

\item  We attempt to simulate the showering that occurs within the ECal. We distribute the energy of each particle that is absorbed in the ECal into a $(3\times 3)$ grid of cells (centered on the direction of the original particle) according to a precomputed Moli\`{e}re matrix corresponding to the Moli\`{e}re radius of lead. For details on this transverse smearing see Appendix~\ref{sec:app-Moliere}.  The structure induced by this shower simulation is observable in our final results.
	 
\item  We implement calorimeter energy smearing for both the ECal and the HCal. The calorimetric response is parametrized through a Gaussian smearing of the accumulated cell energy $E$ with a variance $\sigma$:
\begin{equation}
	\label{eq:cal-response}
	\frac{\sigma}{E} \ = \ \frac{S}{\sqrt{E}} + C  \; ,
\end{equation}
where $S$ and $C$ are the stochastic and constant terms.  For the ECal and the HCal, we use ($S, C$) to be $(0.1,0.01)$ and $(0.5,0.03)$, respectively, in order to approximately match the reported calorimeter response from ATLAS~\cite{Smearing}.  

\item Each calorimeter cell that passes an energy threshold becomes an input for our jet clustering algorithm. For the ECal cells we require $E_T > 0.1~\gev$, while for the HCal cells we use the somewhat harder cut $E_T > 0.5~\gev$.~\footnote{ The specific values are chosen to mimic the choices for real detectors and the difference between the two accounts for the differing noise levels in calorimeter cells of different sizes.}  We sum all the energy deposited  in a given calorimeter cell and construct a massless 4-vector with the 3-vector direction corresponding to the location of that cell.  

          \item As the final step we cluster the 4-vectors corresponding to the calorimeter cells into jets using Fastjet~$3.0.3$~\cite{Cacciari:2005hq, Cacciari:2011ma}. In particular, we use the $\Akt$ jet clustering algorithm~\cite{Cacciari:2008gp} with $R = 0.4$ and and require $p_T > 50~\gev$ for every jet.  Only the leading jet from each event is retained for further analysis in order to maintain independence among the jets in the sample. 
	 
\end{itemize}

\section{\label{sec:analysis} Analysis}

\begin{figure}[h]
	\centering
	\includegraphics[width=0.45\textwidth]{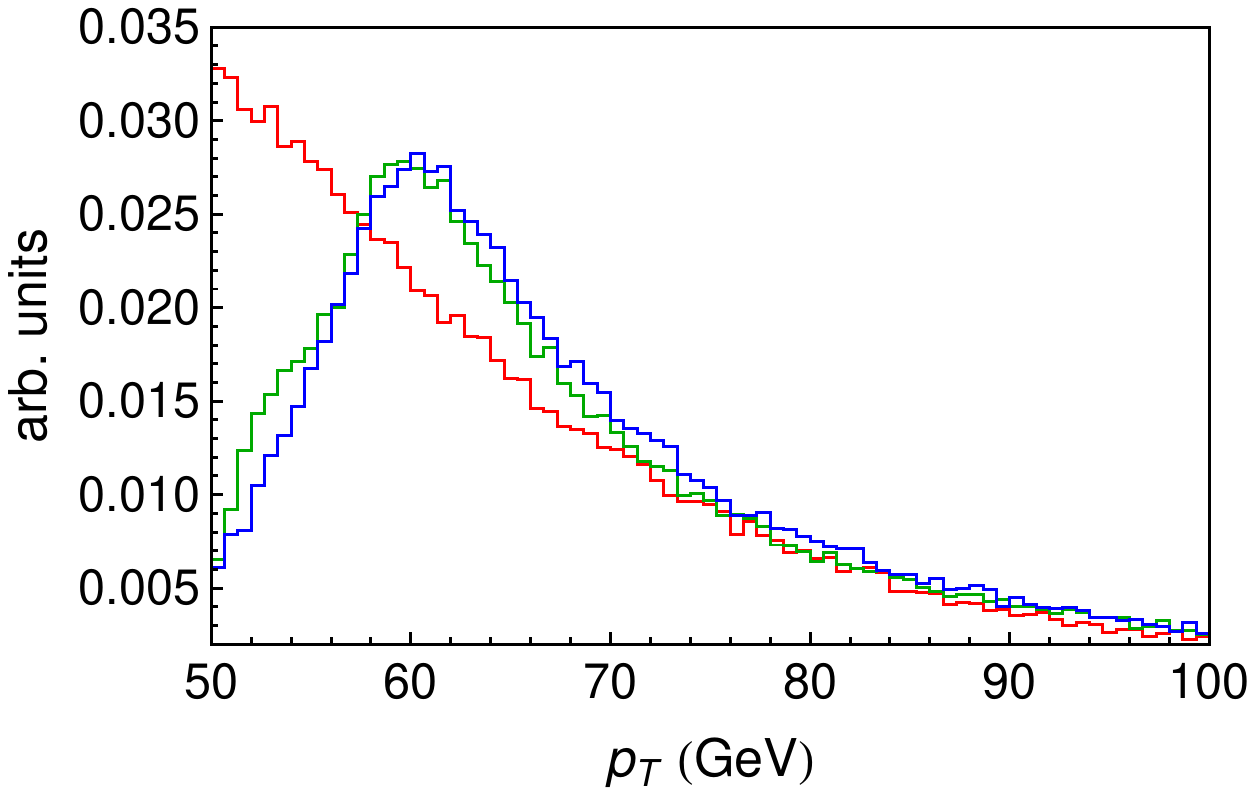}
		\vspace{-3mm}
	\caption{  \label{fig:pT_dist}  The $p_T$ distribution of jets for QCD-jets (red),  single photons (green), and for photon-jets (blue) from the study point \pjsp{8}. Jets are constructed as described in the text (the $\Akt$ algorithm with $R = 0.4$).}
\end{figure}

In this section we describe the analysis of $10$~samples of jets generated according to the prescription of the previous sections. The first sample contains QCD jets derived from QCD dijet events. The second sample consists of jets from $h \rightarrow \gamma \gamma$ events where each jet typically contains one of the photons from the Higgs decays, plus contributions from the rest of the event (ISR, FSR, UE).  We refer to the jets in this sample as single photon jets, or simply single photons.  The remaining $8$ samples of jets are the photon-jet samples and correspond to the $8$ study points in Table~\ref{table:bench}. As noted above, in these events the Higgs particle decays into $4$ or $8$ photons and the corresponding photon-jets typically contain either 2 or 4 photons.  The resulting $p_T$ distributions for QCD-jets (red),  photon-jets (blue) (\pjsp{8}) and single photons (green) are indicated in Fig.~\ref{fig:pT_dist}.  

As expected, the $p_T$ distribution for QCD-jets is a falling distribution, while both the single photon and photon-jet distributions exhibit a peak near $m_h/2 (= 60~\gev)$.   We understand this last point as arising from the production of Higgs particles with zero transverse momentum followed by 2-body decays (either 2 photons or 2 $n_1$'s).  It is the remnants of these two bodies that are typically captured in the jets yielding the indicated peaks near $p_T \sim m_h/2$.  For the photon-jet sample we only show the $p_T$ spectrum for the study point \pjsp{8}, but note that the $p_T$ distributions are almost identical for all other benchmark points.  As indicated in Fig.~\ref{fig:pT_dist}, the jets in all of these samples of events have crudely comparable transverse momentum distributions in the range $50-100~\gev$, although the QCD sample is more strongly peaked at the low end. Thus studying the jets in these samples should provide a useful laboratory in which to study photon-jets, QCD jets and single photons.

The remainder of this section describes a systematic analysis aimed at distinguishing photon-jets from QCD-jets as well as from single photons. We begin with brief descriptions of the variables that provide the discriminating power. The variables are organized into two groups: $(i)$ conventional variables and $(ii)$ substructure variables. We demonstrate how each of these variables individually discriminates photons-jets form the jet samples. Later in this section, we combine these variables in a multivariate analysis in order to maximize the separation of photon-jets  from QCD-jets  as well as from single photons.

\subsection{Conventional Variables}
The conventional variables we describe below are well known, well understood, and play essential roles in the identification of single photons, i.e., the separation from QCD-jets. We expect these variables to play a similar role in separating photon-jets from QCD-jets, since the probability distributions as functions of these variables are similar for photon-jets and for single photons.  On the other hand, they cannot be expected to efficiently discriminate photon-jets from single photons. 

\subsubsection{\label{subsec:theta} Hadronic Energy Fraction, $\theta_J$}

We define the hadronic energy fraction $\theta_J$ for a jet  to be the fraction of its energy deposited in the hadronic calorimeter:
\begin{equation}
 \theta_J  \ = \ \frac{1} {E_J } \sum_{i \in \text{HCal}\, \in \, J} E_{i}
\end{equation}
where $E_{J}$ is the total energy of the jet,  and $E_{i}$ is the energy of the $i$-th HCal cell that is a constituent of the jet.  This is the most powerful variable for discriminating a single photon or a photon-jet (objects that deposit most of their energy in the ECal) from QCD-jets. Since a QCD-jet typically contains 2/3 charged pions and 1/3 neutral pions, we expect to see a peak at $\theta_J \sim 2/3$ ($\log \theta_J \sim -0.2$) for QCD-jets.  Isolated single photons and photon-jets, on the other hand, should exhibit very small $\theta_J$ values.  However, we start with objects identified by a jet algorithm so there will be contributions from the rest of the event and pile-up, and from leakage from the ECal into the HCal.  Thus the precise value of $\theta_J$ for single photons and photon-jets will depend on detailed detector properties and on the contribution from the underlying event and pile-up. Nevertheless, we expect single photons/photon-jets to exhibit very small values of $\theta_j$.

\begin{figure}[h]
\includegraphics[width=0.45\textwidth]{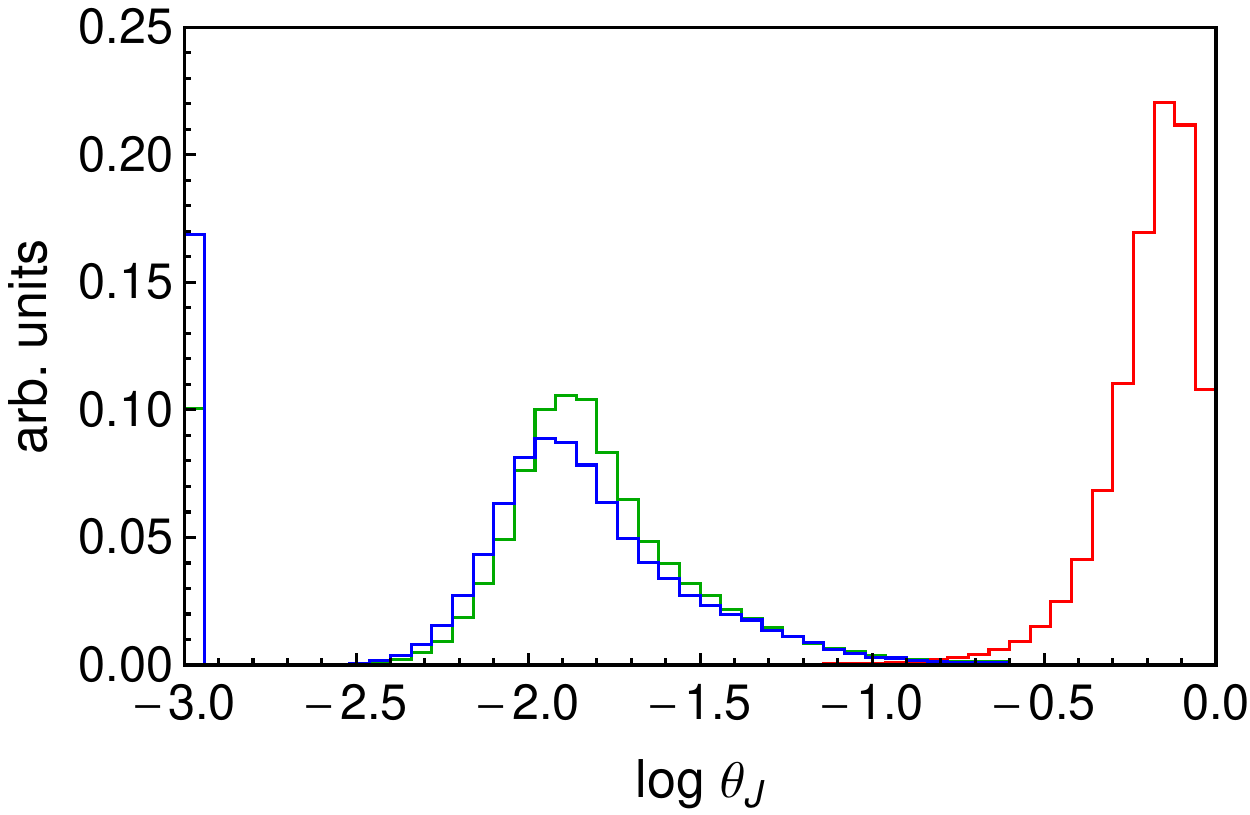}
\caption{\label{fig:hadfrac} The probability distributions for jets as functions of $\log \theta_J$  for QCD-jets (red), single photons (green) and photon-jets from \pjsp{8} (blue).  The first bin of the plot (at $\theta_J = 10^{-3}$) has an open lower boundary, i.e., it includes all jets with  $\log \theta_J < -3.0$.}
\end{figure}

 Figure~\ref{fig:hadfrac} shows the probability distribution versus $\log \theta_J$ for QCD-jets (red), single photons (green), and for photon-jets (blue) in our simulated data. For the photon-jets we only show the study point \pjsp{8}, since the distribution is essentially identical for the other benchmark points.  As expected the QCD-jet distribution peaks near $\log \theta_J = -0.2$ ($\theta_J = 2/3$), while the single photon and photon-jet distributions are very similar with a peak near  $\log \theta_J = -1.9$ and an implied tail to very small $\theta_J$ values.  The clear separation of the single photon/photon-jet distributions from the QCD-jet distribution indicates why this variable plays such an important role in the separation of QCD-jets from photons. 

Any reasonable cut on $\theta_J$ ($\theta_J \sim 0.1$) will reduce the QCD-jet contribution by factors of $10^{-2}$--$10^{-3}$, while barely changing the photon/photon-jet contribution. We impose a preliminary cut by keeping only $\theta_J \leq 0.25$ ($\log \theta_J \leq -0.6$).  About $2\%$ of the original QCD-jets survive this cut, while approximately $94\%$ of the single photons/photon-jets survive. We use the modified jet samples that pass this preliminary $\theta_J$ cut for the remainder of this paper.

\subsubsection{\label{subsec:nu} Number of Charged Tracks, $\nu_J$}

In conventional collider phenomenology, the number of charged particles (tracks) associated with an object is often used to distinguish objects from each other. Although photons and electrons generate similar signatures in the ECal, the latter are typically associated with a track while the former are not. Tracks also play an important role in rejecting QCD-jets since, as mentioned before, a QCD-jet typically contains several charged pions. 

In our simulated data we keep all charged particles with $p_T > 2~\gev$ and assume that all of these correspond to tracks in a real detector.  In order to associate these tracks with the jets, which are constructed entirely from calorimeter cells, we perform the following analysis.  First replace each track by an arbitrarily soft light-like four vector with the same ($\eta$-$\phi$) direction as the track, and then include these soft four-vectors in the jet clustering process along with the calorimeter cells.  (We explicitly check that the inclusion of these soft four-vectors does not affect the outcome of the clustering procedure.)  A track is associated with a jet if the soft four vector corresponding to that track is clustered into that jet%
~\footnote{As a check we also consider the more traditional construction where a track is associated with a jet if it is within an angular distance $R$ or less from the given jet's direction, where $R$ is the size-parameter used in the clustering algorithm.  For $\Akt$ jets both methods yield identical associations of tracks and jets.  For the $\kt$ or $\CA$ algorithms, where jets are not exactly circular, the method described in the text is a more natural definition of whether a track is associated with a jet or not.}. The resulting total number of tracks associated with a jet yields the value of $\nu_J$ for that jet.
\begin{figure}[h]
	\includegraphics[width=0.45\textwidth]{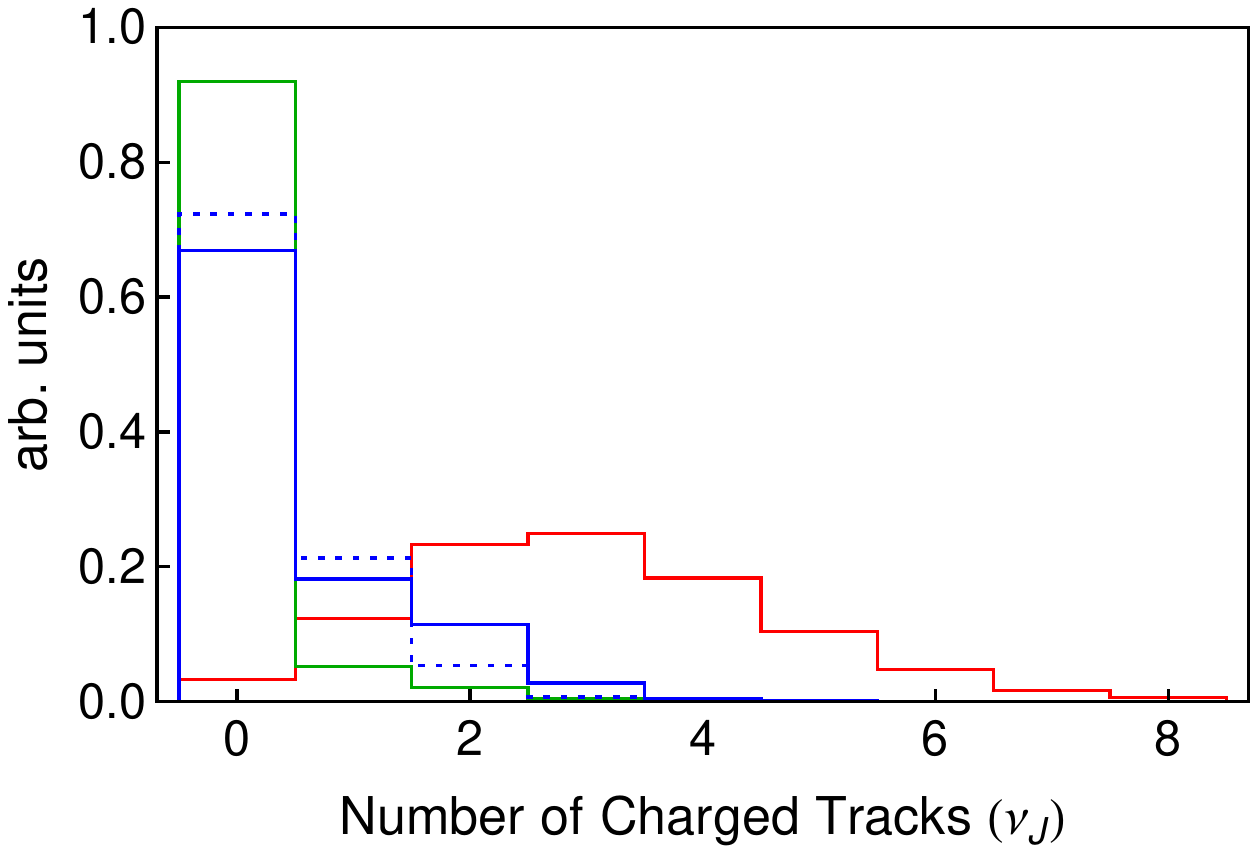}
	\caption{\label{fig:chtrk} The relative probability distribution for QCD jets (red), single photons (green) and photon-jets (blue) versus the number of charged tracks associated with a jet.  The algorithm for associating tracks with jets is given in the text. For photon-jets we show the distribution for jets from the study points \pjsp{1} (dotted) and  \pjsp{8} (solid).  }
\end{figure}
Figure~\ref{fig:chtrk} shows the relative probability distribution versus the number of tracks per jet ($\nu_J$) for QCD-jets (red), single photons (green) and photon-jets (blue).  As expected, the number of tracks associated with QCD-jets varies over a broad range and only a tiny fraction of QCD-jets have no associated tracks. The single photon/photon-jet samples, on the other hand, are dominated by jets with no associated tracks.  Photons that convert yield tracks associated with the corresponding jets. Since the probability of conversion increases with the number of photons per jet, the probability of obtaining one of more associated tracks increases from single photon jets (single photons) to jets with two photons (typical for \pjsp{1}, the dotted blue curve) to jets with four photons (typical for \pjsp{8}, the solid blue curve).  As with the variable $\theta_J$, $\nu_J$ offers some separation between QCD-jets and single photons, but much less between single photons and photon-jets (and even less between the different types of photon-jets).

\subsection{\label{sec:substructure } Jet Substructure}

Next we want to focus on variables that explicitly characterize the internal structure of jets, i.e., characterize the energetic subjet components of the jet.  Recall that in this analysis we have identified jets using the the $\Akt$ jet algorithm with $R=0.4$, but we do not expect the general features of our analysis to depend on this specific choice.  The next step is to determine a `recombination tree' for the jets we want to study (here the leading jet in each event).  To this end we apply the $\kt$ algorithm~\cite{Catani:1993hr, Ellis:1993tq} to the calorimeter cells identified as constituents of the jet in the first step. (We could as well use the Cambridge/Aachen ($\CA$) algorithm~\cite{Dokshitzer:1997in, Wobisch:1998wt, Wobisch:2000dk}, but not the $\Akt$ algorithm in this step as $\Akt$ does not tend to produce a physically relevant recombination tree.)   This recombination tree specifies the subjets at each level of recombination $N$ from $N=1$ (the full jet) to $N=$ the number of constituent calorimeter cells in the jet (no recombination).  At the next step the subjet variables we study fall into two classes.  In the first class we attempt to count the effective number of relevant subjets without using any properties of the subjets in the tree except their directions in $\eta$-$\phi$.  In this case the useful variable (defined in detail below) is called $N$-subjettiness.  The $N$-subjettiness variable for a given jet becomes numerically small when the parameter $N$ is large enough to describe all of the relevant substructure, i.e., this value of $N$ provides a measure of the number of subjets without explicitly identifying the subjets. $N$-subjettiness involves \textit{all} components of the original jet for all values of $N$.

The rest of the substructure variables we study more explicitly resolve a jet into a set of subjets.  We define both the level in the recombination tree at which we choose to work, i.e., the number of subjets we have split the jet into and how many of these subjets to use in the subsequent analysis.  We use $N_\text{pre-filter}$ (this notation should become clear shortly) and $N_\text{hard}$ to label these two parameters.  Thus we start with the 4-vectors corresponding to the (calorimeter cell) constituents of a given jet, and then (re)cluster these constituents using the chosen subjet algorithm (which is not necessarily the algorithm used to originally identify the jet) in \textit{exclusive} mode, i.e. we continue (re)clustering until there are precisely $N_\text{pre-filter}$ 4-vectors left -- the $N_\text{pre-filter}$ exclusive subjets. Out of these  $N_{\text{pre-filter}}$ subjets we pick the $N_\text{hard}$ largest $p_T$ subjets and discard the rest. All the substructure variables discussed below (except  $N$-subjettiness) are constructed using these $N_\text{hard}$ subjets.  Note that by choosing $N_{\text{pre-filter}} > N_\text{hard}$, we have performed a version of jet `grooming' typically labeled filtering~\cite{Butterworth:2008iy, Butterworth:2008sd, Butterworth:2008tr}.  This will ensure that our results are relatively insensitive to the effects of the underlying event and pile-up.    Ideally,  the integers $(N_\text{hard}, N_{\text{pre-filter}})$ should be chosen based on the topology of the object we are looking for. However, the naive topology will be influenced by the interaction with the detector and the details of the jet clustering algorithm.  For example, 
a $4$ photon photon-jet will often appear in the detector to have fewer than $4$ distinct lobes of energy, i.e., one or more photons often merge inside a single lobe of energy.   In our simulation, we find that the choice $N_\text{hard} = 3$  and $N_\text{pre-filter} = 5$ is an acceptable compromise, working reasonably well for single photons and photon-jets from all the study points.  Further optimization will be possible in the context of real detectors and searches for specific photon-jet scenarios.

\subsubsection{\label{subsec:tau}  $N$-Subjettiness, $\tau_N$}

\begin{figure*}[t]
\centering
	\subfloat[]{\includegraphics[width=0.25\textwidth]{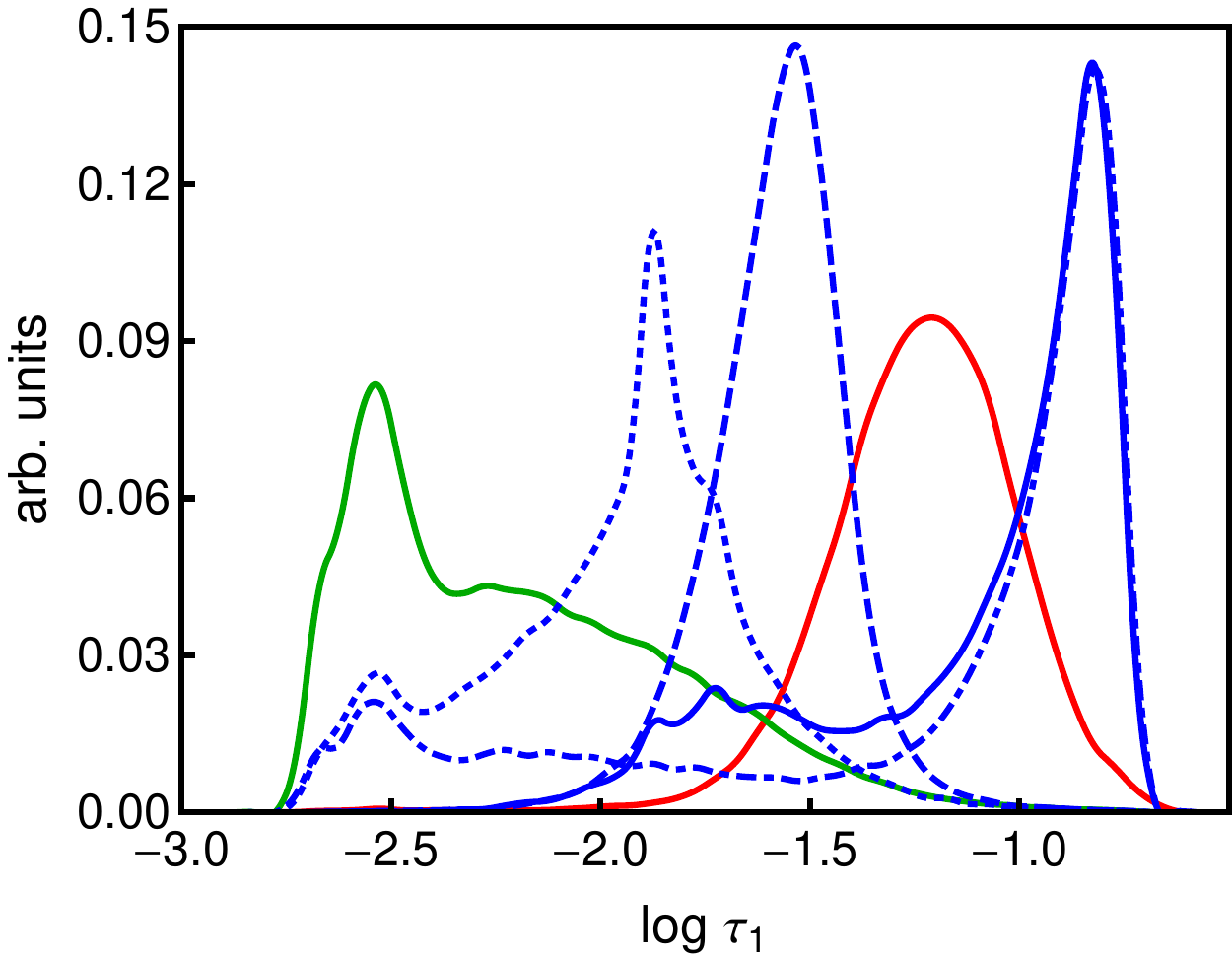}}  \hspace{-1mm}
	\subfloat[]{\includegraphics[width=0.25\textwidth]{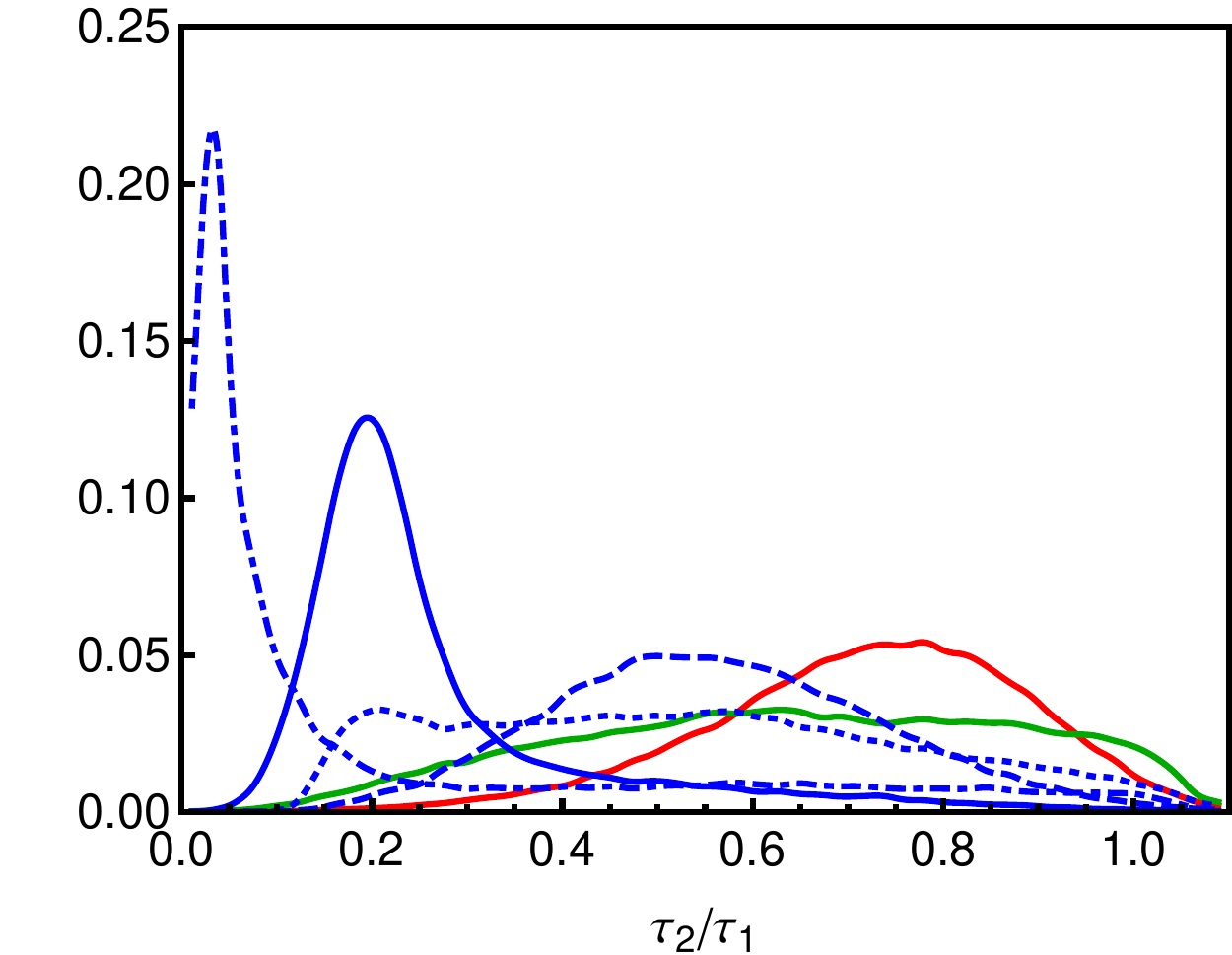}} \hspace{-1mm}
	\subfloat[]{\includegraphics[width=0.25\textwidth]{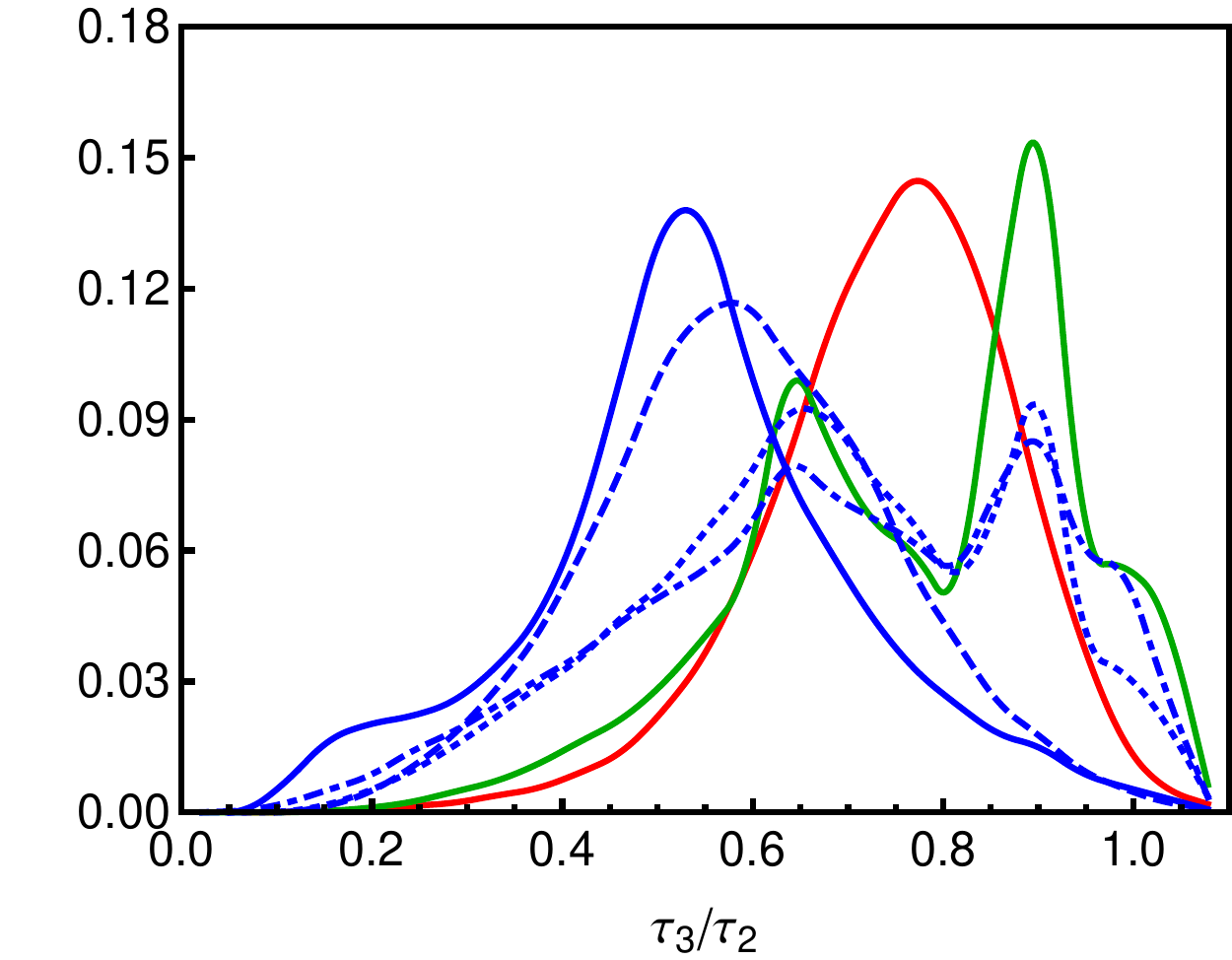}} \hspace{-1mm}
	\subfloat[]{\includegraphics[width=0.25\textwidth] {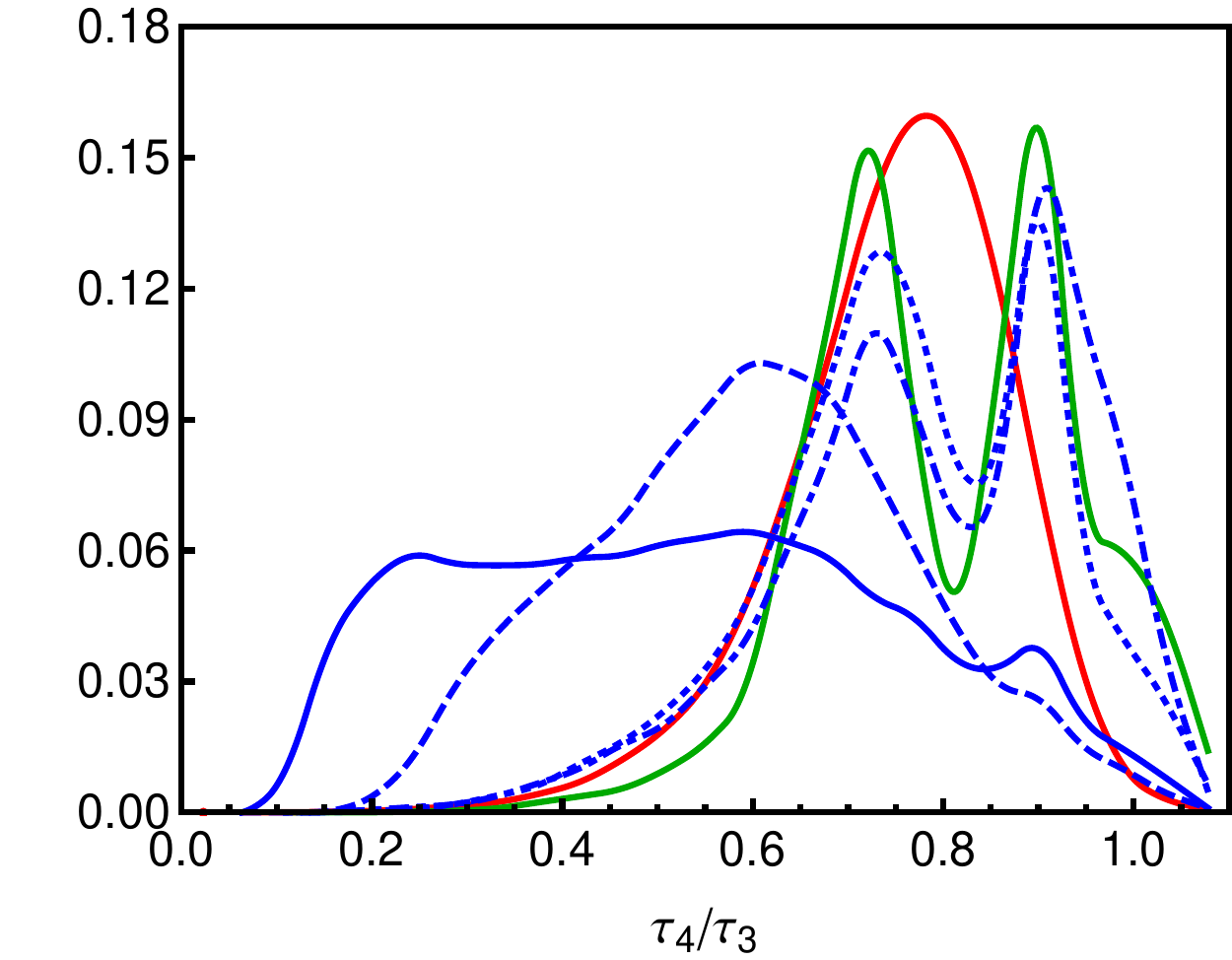}}
		\vspace{-5 mm}
\caption{\label{fig:tau} Probability distributions vesus various $N$-subjettiness variables. The solid red and green curves show, as usual, the distributions for QCD-jets  and single photons respectively. Various blue curves are for photon-jets from different study points. The solid, dashed, dotted and dash-dotted curves in all these figures are for \pjsp{8}, \pjsp{4}, \pjsp{1} and \pjsp{3} respectively.}
\end{figure*}

``$N$-subjettiness", introduced in Ref.~\cite{Thaler:2010tr, Thaler:2011gf}, is a modified version of  ``$N$-jettiness"  from Ref.~\cite{Stewart:2010tn}. It is adapted in a way such that it becomes a property of a jet rather than of an event. $N$-subjettiness provides a simple way to effectively count the number of subjets inside a given jet.  It captures 
whether the energy flow inside a jet deviates from the one-lobe configuration expected to characterize a typical QCD-jet.  We use the definition of $N$-subjettiness proposed in Ref.~\cite{Thaler:2010tr}.  The starting point is a jet, the full set of  4-vectors corresponding to the (calorimeter cell) constituents of the jet (here found with the $\Akt$ algorithm for $R=0.4$), and the recombination tree found with the $\kt$ algorithm as outlined above.  From this tree we know the 4-vectors describing the exclusive subjets for any level $N$, i.e., the level where there are exactly $N$ subjets.  With this information we can define $N$-subjettiness to be   
\begin{equation}
	\label{eq:tau}
	\tau_N = \frac{ \sum_k  p_{T_{k}}  \times \text{min} \bigl \{ \Delta R_{1,k}, \Delta R_{1,k} , 
				\cdots, \Delta R_{N,k}  \bigr \} } { \sum_k  p_{T_{k}} \times R}  \; ,
\end{equation}
where $k$ runs over all the (calorimeter cell) constituents of the jet, $p_{T_{k}}$ is the transverse momentum for the $k$-th constituent, $\Delta  R_{l,k}=\sqrt{(\Delta \eta_{l,k})^2+ (\Delta \phi_{l,k})^2} $ is the angular distance between the $l$-th subjet (at the level when there are $N$ subjets) and the $k$-th constituent of the jet, and  $R$  is the characteristic jet radius used in the original jet clustering algorithm.

In the context of single photons, photon-jets and QCD-jets, we use $N$-subjettiness in two different ways. The first application is to use the \textit{ratios} $\tau_{N+1}/ \tau_{N}$ in the same way $N$-subjettiness is used to tag boosted massive particles such as a $W$~boson or a hadronic decaying top~\cite{Thaler:2010tr, Thaler:2011gf}.  In particular, for a jet with $N_0$ distinct lobes of energy, $\tau_{N_{0} } $ is expected to be much smaller than $\tau_{N_{0} -1} $ (of course, we are assuming $N_0 > 1$), whereas for $N > N_0$,  $\tau_{N+1}$ is expected to be comparable to $\tau_{N}$.   Thus a two photon photon-jet is expected to be characterized by $\tau_2/\tau_1 \ll 1$.  On the other hand, one lobed QCD-jets and single photons should exhibit comparable values for $\tau_2$ and $\tau_1$, and consequently $\tau_2/\tau_1 \sim 1$. 

The second  way in which we use $N$-subjettiness consists of using the magnitude of $\tau_1$ itself.  Even for a jet with one lobe of energy the exact magnitude of $\tau_1$ represents a measure of how widely the energy is spread. A pencil-like energy profile, like that of a single photon or a narrow photon-jet, should yield a much smaller $\tau_1$ compared to QCD-jets with a much broader profile. In fact, $\tau_1$ is an indicator of jet mass, and, for jets with identical energy, $\tau_1$ is proportional to the square of the jet mass.

Figure~\ref{fig:tau} shows the probability distributions versus $\log \tau_1$ and $\tau_{N+1}/ \tau_{N}$  for $N = 1,2,3$ corresponding to single photons, QCD-jets and photon-jets from different study points. Note that for photon-jets, the jet mass is almost always given by the mass parameter $m_1$  in Table~\ref{table:bench}. Thus for \pjsp{8} and \pjsp{3}, where $m_1$ has the same value, the probability distributions versus $\log \tau_1$ are almost identical.  For study points \pjsp{8}, \pjsp{4} and \pjsp{1} the peak in $\log \tau_1$ shifts to the left as the value of $m_1$ decreases (from $10~\gev$ to $2~\gev$ to $0.5~\gev$). Note also that the \pjsp{1} and \pjsp{3} distributions exhibit a small $\tau_1$ (small mass) enhancement at essentially the same $\tau_1$ value as the primary peak in the single photon (green curve) distribution. This presumably corresponds to those kinematic configurations where only one of the (two) photons from the $n_1$ decay is included in the jet. Thus we expect that a (small) fraction of the time these scenarios will look very single photon-like.

Clearly the ratio $\tau_2/\tau_1$ gives significant separation for the different photons-jet scenarios.  The study points \pjsp{8} and \pjsp{3} are now separated, although both exhibit peaks at small values of the ratio.  This suggests an intrinsic 2-lobe structure corresponding to 2 photons for \pjsp{3} and 4 photons in two relatively tight pairs ($m_2 \ll m_1$) for  \pjsp{8}.  \pjsp{4} with presumably a more distinctive 4 photon structure exhibits a broader peak at a larger value of $\tau_2/\tau_1$.  Single photons and  \pjsp{1} exhibit even broader distributions presumably corresponding to an intrinsically 1-lobe structure.  The QCD-jet distribution is also broad but with an enhancement around $\tau_2/\tau_1 = 0.8$, presumably arising from a typical 1-lobe structure but some contribution from showers with more structure and from the underlying event.  The ratios $\tau_3/\tau_2$ or $\tau_4/\tau_3$ seem to be less effective in discriminating photon-jets from single photons and QCD-jets. This can be understood by noting that quite often the hard photons inside a photon-jet become collinear at the scale of the size of the cell. So even for photon-jets with  $4$ hard photons, we rarely find jets with  $4$  distinct centers of energy.  In general we expect the ratio $\tau_{N+1}/ \tau_{N}$  becomes less and less useful with increasing $N$.  

Note that the distributions for single photons and photon-like photon-jets tend to exhibit a double peak structure in $\tau_3/\tau_2$ or $\tau_4/\tau_3$.  We believe that this feature arises from both the contributions due to the underlying event and due to our implementation of transverse smearing in the ECal (see Appendix~\ref{sec:app-Moliere}). 

\subsubsection{\label{subsec:lambda} Transverse momentum of the Leading Subjet}

Now we proceed to discuss the second class of subjet variables constructed from the 3 hardest subjets out of the 5 exclusive subjets.  As the first such variable consider the fraction of the jet transverse momentum carried by the leading subjet, which provides significant information about the jet itself. In particular, it indicates the fraction of the jet's total $p_T$ carried by the leading subjet only. Since photon-jets result from the decay of massive particles into hard and often widely separated photons inside the jet, the subjets are usually of comparable hardness.  The leading subjet for single photons and for QCD-jets, on the other hand, typically carry nearly the entire $p_{T}$ of the jet. So for the majority of these jets, the $p_{T}$ of the leading subjet (label it $p_{T_{L}}$)  is of the order of the $p_{T}$ of the entire jet ($p_{T_{J}}$).  Instead of using the ratio $p_{T_{L}} / p_{T_{J}}$ directly we find that it is more instructive to define the variable
\begin{equation}
	\label{eq:lambda}
	\lambda_J \ = \ \log \Bigl(1- \frac{p_{T_{L}}} {p_{T_{J}}} \Bigr) \; .
\end{equation} 
The advantage of using the definition in Eq.\eqref{eq:lambda} is that it focuses on the behavior near $p_{T_{L}} \sim p_{T_{J}}$.  

The discussion above depends crucially on how the subjets are constructed, especially for QCD-jets.  QCD partons typically shower into many soft partons/hadrons.  After showering and hadronization, single hard partons yield many soft hadrons distributed throughout the jet.  The way in which these jets are clustered into subjets dictates the $p_T$ distribution of subjets.  For example, for $\Akt$ subjets, the hardest subjet will always have $p_{T_{L}} \simeq p_{T_{J}}$.  The $\kt$ algorithm, on the other hand, clusters the softer elements first and results in more evenly distributed subjets. The $\CA$ jet algorithm clusters taking into consideration only the geometric separations of the elements, and produces qualitatively different results. Single photons, on the other hand, shower very little (no QCD Shower) and deposit energy in only a handful of cells (per hard photon). Therefore we expect that our results for single photons or photon-jets will be less sensitive to the details of the clustering algorithm.   To versify this point we use both $\kt$  and $\CA$ subjets
to evaluate $\lambda_J$ from Eq.\eqref{eq:lambda}.  The simultaneous use of different clustering  algorithms to extract information from the same jet should not come as a surprise. As shown in  Ref.~\cite{Ellis:2012sn}, substantial further information  can be extracted if one employs a broad sampling out of \textit{all} of the physically sensible clustering histories (trees) for a given jet.  In this sense the current analysis is  modest in that  we only use two specific clustering procedures.

\begin{figure}[h]
\centering
	\subfloat[]{\includegraphics[width=0.24\textwidth]{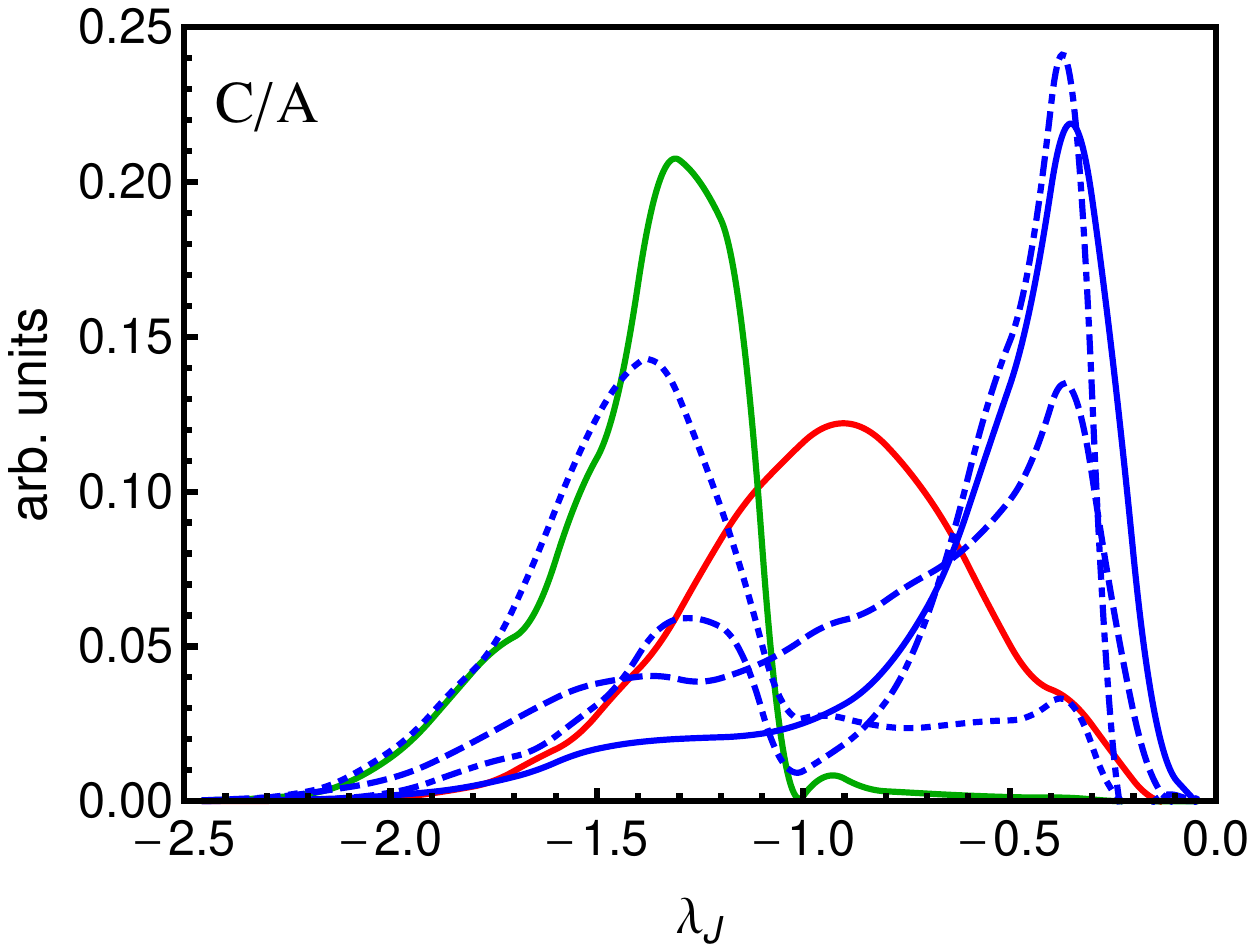}} \hspace{-1mm}
	\subfloat[]{\includegraphics[width=0.24\textwidth]{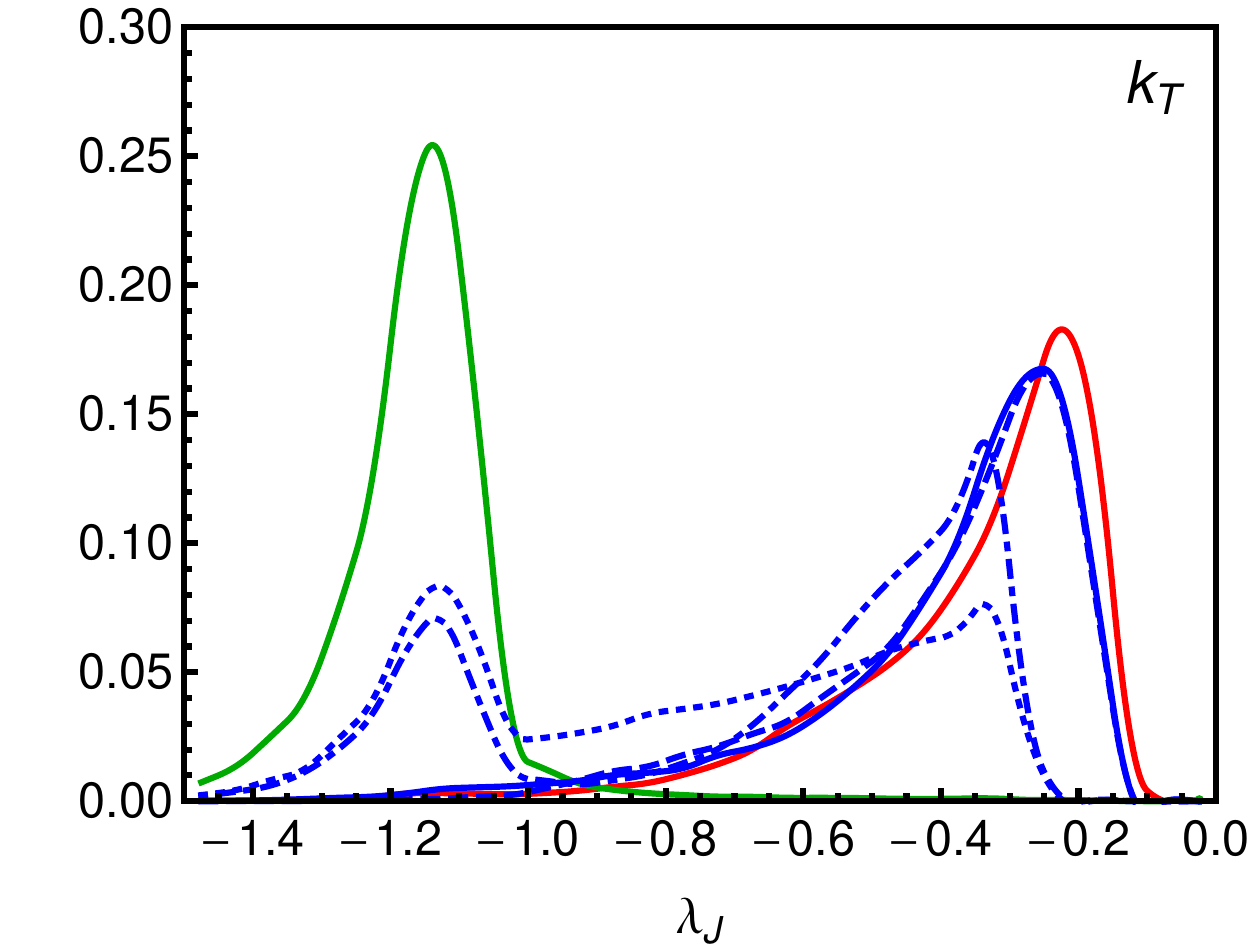}}
		\vspace{-5 mm}
\caption{\label{fig:lambda} Probability distribution for $\lambda_J$ from Eq.\eqref{eq:lambda}. As in Fig.~\ref{fig:tau} the solid red is for QCD-jets, the solid green for single photons, dotted blue for \pjsp{1}, dash-dotted blue for \pjsp{3}, dashed blue for  \pjsp{4} and solid blue for \pjsp{8}.  The left (right) figure shows the distribution when $\lambda$ is calculated using $\CA$ ($\kt$) subjets.}
\end{figure}
In Fig.~\ref{fig:lambda} we  plot the probability distribution of jets as a function of $\lambda_J$ for QCD-jets, 
single photons, and photon-jets. The left (right) panel shows the distribution when we use the $\CA$ ($\kt$) algorithm to find the subjets.  Note how the distribution for QCD-jets (the red curve) moves more to the right (i.e., the $p_{T}$ of the jet gets more evenly distributed among its subjets) as we go from $\CA$ subjets to $\kt$ subjets.  The various photon-jet study points also look more similar when using the $\kt$ algorithm.  In this case the \pjsp{1} and \pjsp{3} distributions exhibit enhancements suggesting the presence of both single photon-like behavior ($\lambda_J \sim -1.2$) and QCD-like behavior ($\lambda_J \sim -0.2$ to $-0.3$).  The more complex structure of the \pjsp{4} and \pjsp{8} jets exhibit a distribution closer to QCD alone.  Finally note that the $\CA$ subjets display the jet substructure information differently from the $\kt$ case with the peak in the QCD-jet distribution at least somewhat separated from the peaks in the photon-jet distributions. Also for $\CA$ all of the photon-jet scenarios exhibit at least a little single photon-like enhancement (for $\kt$ this is  only true for \pjsp{1} and \pjsp{3}).   There is clearly some discrimination to be gained from using more than one definition of the subjets.

\subsubsection{\label{subsec:epsilon} Energy-Energy Correlation, $\epsilon_J$ } 

Another useful variable is the ``energy-energy correlation". We define it as:
\begin{equation}
	\label{eq:epsilon}
 	\epsilon_J \ = \  \frac{1}{E_J^2}  \sum_{(i > j) \in N_\text{hard} }  E_{i} E_{j}\,,
\end{equation}
where $E_J$ is the total energy of a given jet, and the indices $i,j$ run over the (3 hardest) subjets of the jet. From the definition, it should be clear that $\epsilon_J$  is sensitive to the energy of the subleading jets. In particular, the energy-energy correlation can be expressed as
\begin{eqnarray}
\epsilon_J  & = \frac{ E_{L}\left(E_{NL}+ E_{NNL} \right) + E_{NL}  E_{NNL}}{E_J^2}\nonumber\\
                  & \approx  \frac{ E_{L}\left(E_{J}- E_{L} \right) + E_{NL}  E_{NNL}}{E_J^2} \,, 
\end{eqnarray}
where $E_{L},E_{NL}$, and $E_{NNL}$ are the energies of the leading subjet, the next-to-leading subjet, and the next-to-next-to-leading subjet. 

\begin{figure}[h]
\centering
	\subfloat[]{\includegraphics[width=0.24\textwidth]{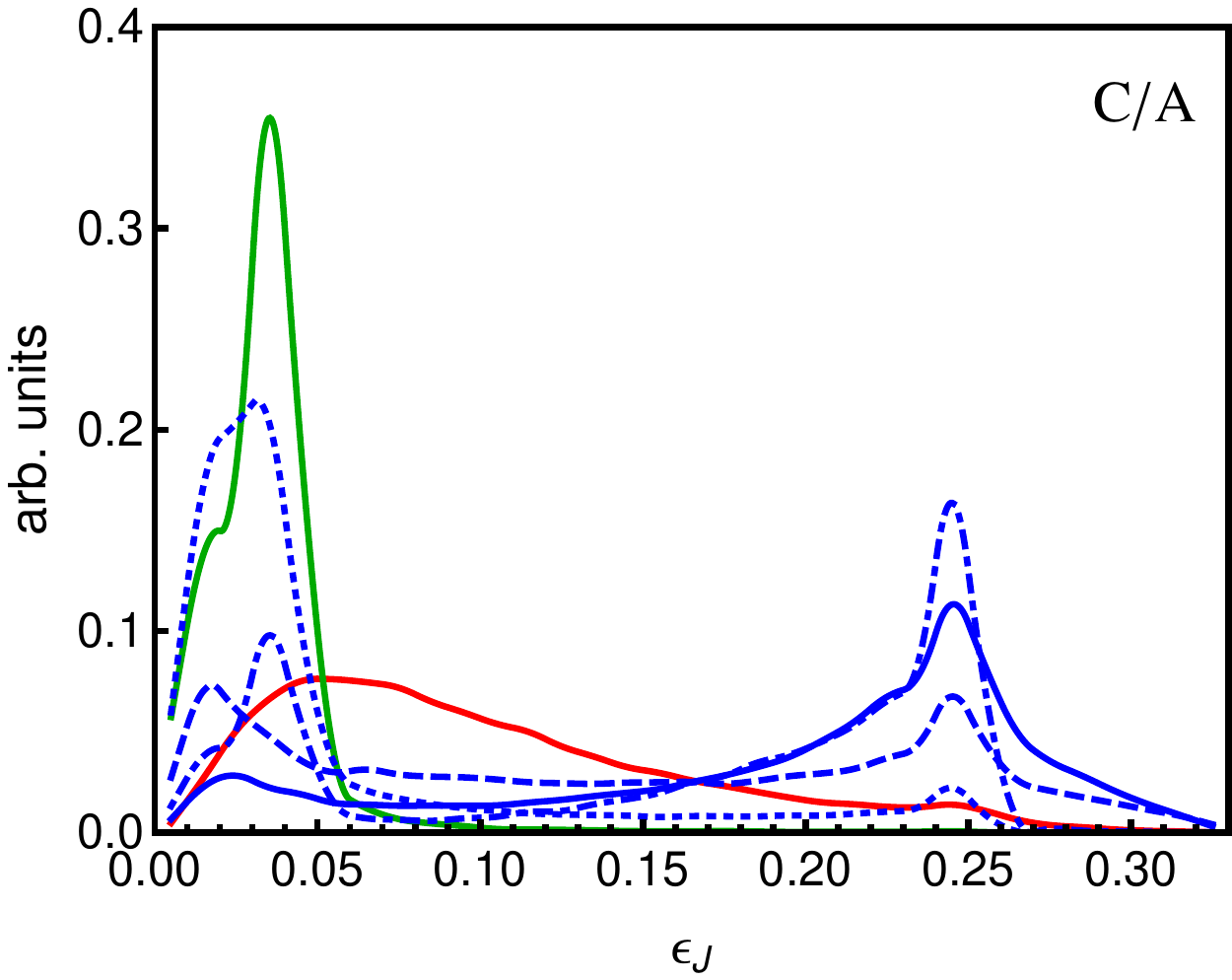}} \hspace{-1mm}
	\subfloat[]{\includegraphics[width=0.24\textwidth]{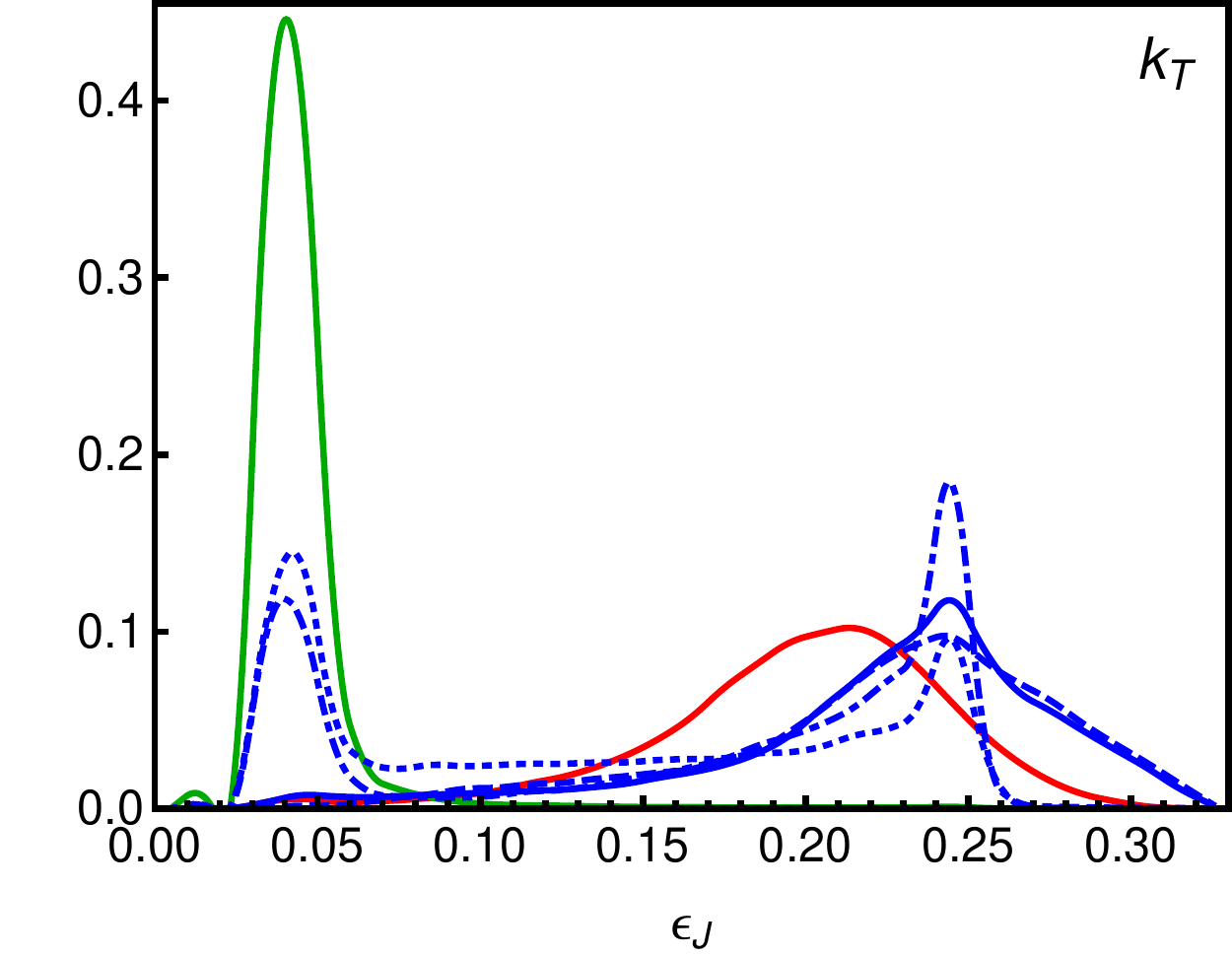}}
		\vspace{-5 mm}
\caption{\label{fig:epsilon} Probability distribution versus $\epsilon_J$ from Eq.\eqref{eq:epsilon}. As in Fig.~\ref{fig:tau} the solid red is for QCD-jets, the solid green for single photons, dotted blue for \pjsp{1}, dash-dotted blue for \pjsp{3}, dashed blue for  \pjsp{4} and solid blue for \pjsp{8}.  The left (right) figure shows the distribution when $\epsilon_J$ is evaluated using $\CA$ ($\kt$) subjets.}
\end{figure}
We show the probability distribution of jets as a function of $\epsilon_J$ for QCD-jets, single photons and photon-jets in Fig.~\ref{fig:epsilon}. Note that for single photons (the green curve), $E_{NL} \text{ and } E_{NNL}$ are negligible and hence we expect $\epsilon_J$ for single photons to be well approximated by $E_{L}\left(E_J- E_{L} \right)  /E_J^2 $. In fact, the sharp peak for single photons in Fig.~\ref{fig:lambda} at $-1.2$ ($\kt$ algorithm) corresponds to the sharp peak at about $0.04$ in Fig.~\ref{fig:epsilon}.  More generally the qualitative features in  Fig.~\ref{fig:lambda} are repeated in  Fig.~\ref{fig:epsilon}.  For $\CA$ subjets the distributions for all of the photon-jet study points exhibit two peaks, the large $\epsilon_J$ value enhancement presumably corresponding to the energy being shared approximately equally among several final photons, while the small value enhancement arises from the case when one photon dominates (perhaps because some of the photons are not in the jet).  For $\kt$ subjets only the \pjsp{1} and \pjsp{3} distributions exhibit the small $\epsilon_J$ single photon-like enhancement.  We also see that again the two algorithms yield distinctly different distributions for QCD-jets.

\subsubsection{\label{subsec:rho} Subjet Spread, $\rho_J$}

We define ``subjet spread" as a measure of the geometric distribution of the subjets.
\begin{equation}
	\label{eq:rho}
	 \rho_J = \frac{1}{R} \sum_{(i > j) \in N_\text{hard}}  \Delta R_{i,j} \; ,
\end{equation}
where $ \Delta R_{i,j}$ is the angular distance between the $i$-th and $j$-th (hard) subjets, and $R$ is the size parameter of the jet algorithm.

\begin{figure}[h]
	\subfloat[]{\includegraphics[width=0.24\textwidth]{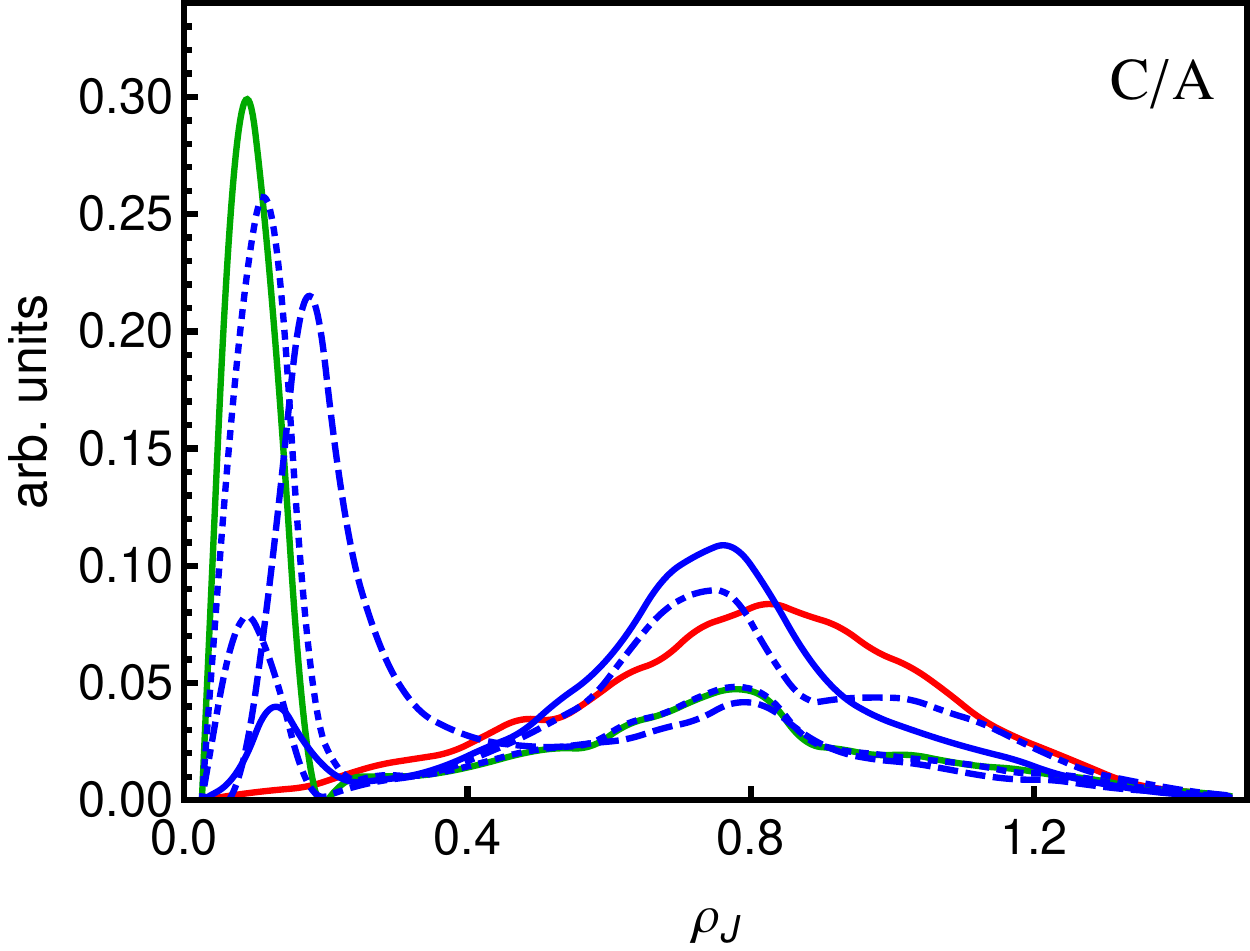}} \hspace{-1.mm}
	\subfloat[]{\includegraphics[width=0.24\textwidth]{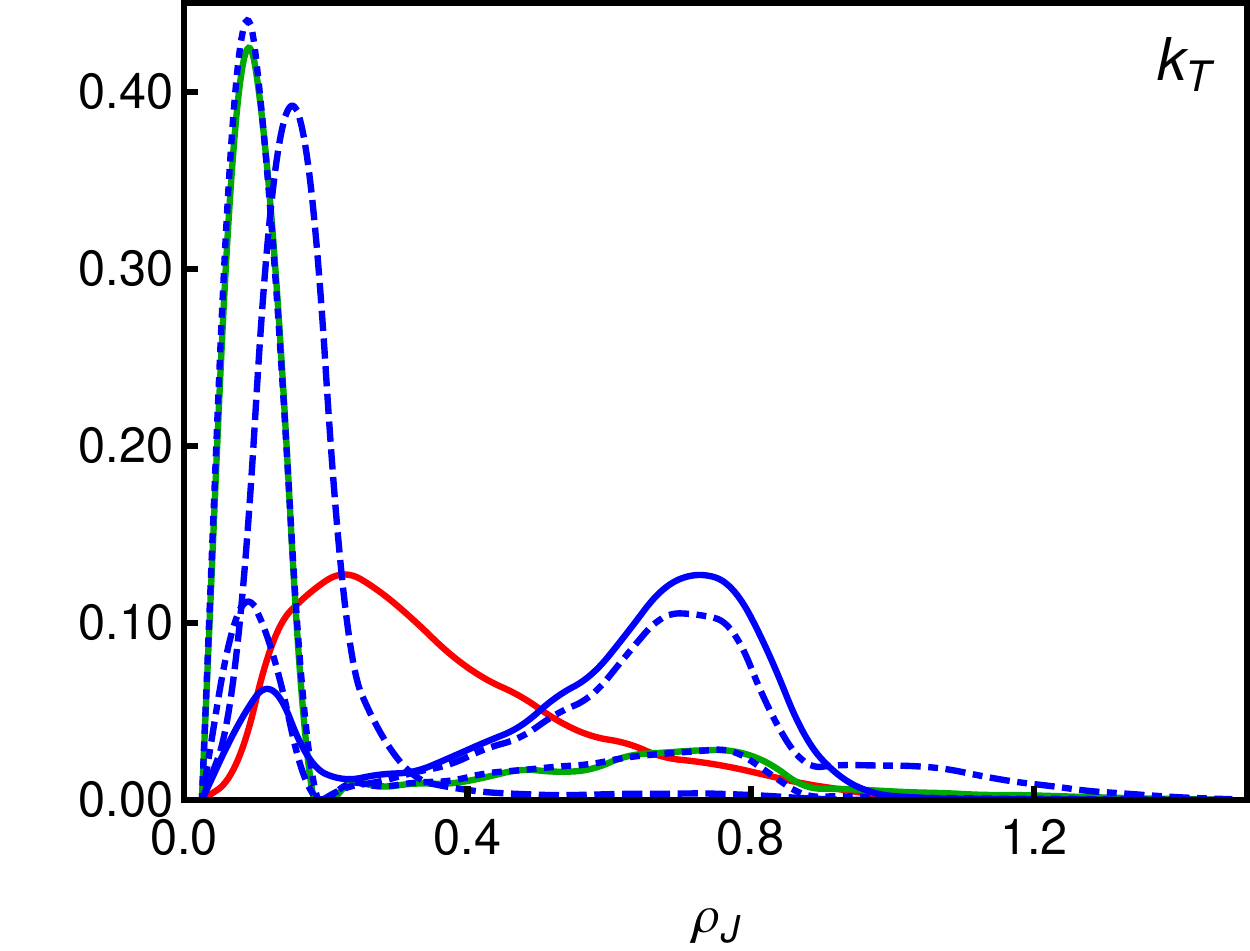}}
		\vspace{-5 mm}
\caption{\label{fig:rho} Probability distribution for subjet-spread $\rho_J$ from Eq.\eqref{eq:rho}. As in Fig.~\ref{fig:tau} the solid red is for QCD-jets, the solid green for single photons, dotted blue for \pjsp{1}, dash-dotted blue for \pjsp{3}, dashed blue for  \pjsp{4} and solid blue for \pjsp{8}. The left (right) figure shows the distribution when $\rho_J$ is calculated using $\CA$ ($\kt$) subjets.}
\end{figure}
The left (right) panel of Fig.~\ref{fig:rho} shows the probability distribution of jets as a function of $\rho_J$ for QCD-jets, single photons and photon-jets when the $\CA$ ($\kt$) subjets are used to evaluate Eq.\eqref{eq:rho}. 
For this variable only the QCD-jet distribution changes dramatically when changing the choice of subjet algorithm from $\CA$ to $\kt$.  By using both algorithms this feature will provide some ability to discriminate between QCD-jets and single photons or photon-jets.  For the single photon case the the strong peak at small $\rho_J$ confirms that all of the subjets are close to each other, forming a hard core. Subjet spread is quite sensitive to the mass $m_1$ as can be seen from the different photon-jet distributions.  In particular, the position of the peaks for photon-jets with different $m_1$ simply follow the $m_1$ value.  The \pjsp{3} and \pjsp{8} distributions are nearly the same (with the same $m_1$ value), while the \pjsp{1} and \pjsp{4} distributions are just similar (with somewhat different $m_1$ values), but distinct from \pjsp{3} and \pjsp{8}. The $m_1$ dependence is not surprising since the opening angle between the decay products of the $n_1$ particle depends on  $m_1$.  Finally we note that the \pjsp{3} and \pjsp{8} distributions do have an enhancement at small $\rho_J$ values presumably corresponding to configurations where the extra photons are not captured in the jet.

\subsubsection{\label{subsec:delta} Subjet Area of the Jet}

As defined in Ref.~\cite{Cacciari:2008gn}, the ``area" associated with a jet is an unambiguous concept that represents quantitatively the amount of surface in the ($\eta$-$\phi$) plane included in a jet.  In this analysis, we use the ``active area" definition for the area of the jet. The active area of a jet is calculated by adding a \textit{uniform} background of arbitrarily soft `ghost' particles to the event (so that each ghost represents a fixed area) and then counting the number of ghosts clustered into the given jet.
The area of a jet is often used to provide a quantitative understanding of the largely uncorrelated contributions to a jet from the underlying event and pile-up.  However, it is rarely used in phenomenology for the purpose of discovering new particles or tagging jets.  We use `subjet area' as a measure of the `cleanliness' of the  jet.  We show that it can be a useful tool for distinguishing a single photon or a photon-jet from noisier QCD-jets. We define the subjet area fraction as
\begin{equation}
	\label{eq:delta}
	\delta_J = \frac{1}{A_J} \sum_{i \in N_\text{hard}} A_i  \; ,
\end{equation}
where $A_i$ is the area of the $i$-th subjet and $A_J$ is the area of the entire jet.  Note that this definition of $\delta_J$ is only useful when the subjets are constructed geometrically by merging the nearest neighbors first (i.e., using the $\CA$  algorithm).   
\begin{figure}[h]
	\includegraphics[width=0.45\textwidth]{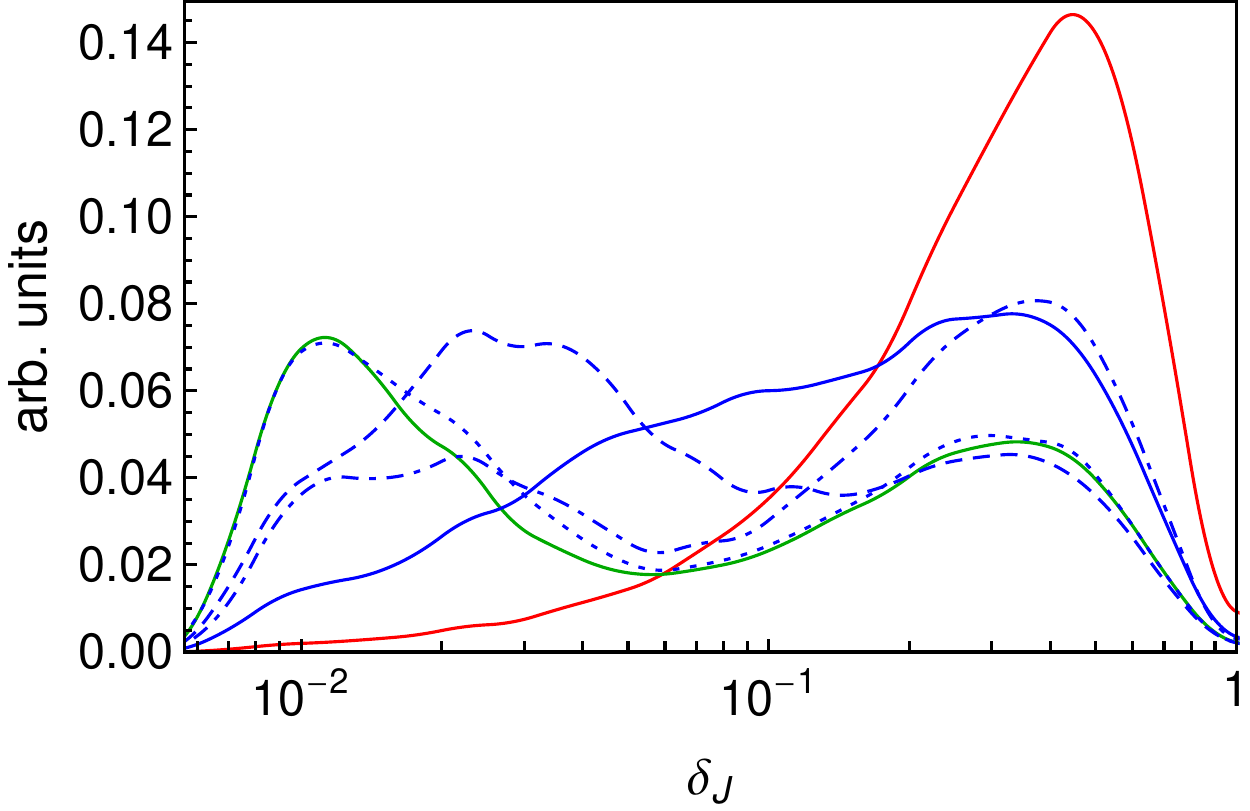}
		\vspace{-3mm}
\caption{\label{fig:delta} Probability distribution versus fractional area $\delta_J$ from Eq.\eqref{eq:delta}. As in Fig.~\ref{fig:tau} the solid red is for QCD-jets, the solid green for single photons, dotted blue for \pjsp{1}, dash-dotted blue for \pjsp{3}, dashed blue for \pjsp{4} and solid blue for \pjsp{8}.  We use $\CA$ subjets to calculate $\delta_J$.}
\end{figure}
In Fig.~\ref{fig:delta}, we show the probability distribution for jets as a function of  $\delta_J$ for QCD-jets, single photons, and photon-jets. As expected, the figure shows that single photons (the green curve) are significantly cleaner (exhibit smaller $\delta_J$ values) than QCD-jets (the red curve) and  that photon-jets (the blue curves) tend to lie in between.  Fixing $m_1$ such that the first splitting is fairly wide, we can investigate the effects of $m_2$. If $m_2$ is small, then the two photons coming from the $n_2$ decays will be very close together, and the subjet that contains them will not collect many ghosts. On the other hand, a large $m_2$ will split the two photons further apart and, if still contained in the same subjet, that subjet will collect substantially more ghosts resulting in a subjet with a larger active area. QCD-jets contain many soft particles and so the subjets in QCD jets have larger areas.  Thus we see that the QCD distribution peaks for $\delta_j$ near $0.5$, while the single photon distribution exhibits both a large peak at small ($\sim 10^{-2}$) $\delta_J$ and a smaller peak at larger ($\sim 0.4$)  $\delta_J$ values.  The photon-jet cases interpolate between these two behaviors and this variable can clearly provide some discriminating power.

\subsection{\label{sec:results} Multivariate Analysis}

We have, so far, introduced a set of well-understood variables.  In this subsection, we will employ these variables in a multivariate discriminant, specifically in a Boosted Decision Tree (BDT)~\cite{BDT}.  A decision tree is a hierarchical set of one-sided cuts used to discriminate signal versus background.  The `boosting' of a decision  tree extends this concept from one tree to several trees which form a forest. The trees are derived from the same training ensemble by reweighing events, and are finally combined into a single classifier.  

In the current discussion we are treating photon-jets as the signal and both single photons and QCD-jets as background. We construct multiple BDT analyses in order to estimate how well the photon-jets can be separated from single photons \textit{and} from QCD-jets.  This will allow us to demonstrate the power of the new jet substructure variables when these are combined with  the conventional variables.  In practice, we employ the Toolkit for Multivariate Analysis (TMVA)~\cite{Hocker:2007ht} package and use the ``BDTD" option to book BDTs, where the input variables are decorrelated first. 

For every study point in Table~\ref{table:bench} we optimize two separate BDTs, one for discriminating photon-jets form QCD-jets and the other for separating photon-jets from single photons. We make use of all the variables discussed earlier in order to minimize the background fake rate ($\mathcal{F} =$ the fraction of the background jets that pass the cuts) for a given signal acceptance rate ($\mathcal{A} = $ the fraction of the signal jets that pass the cuts).  For demonstration purposes we also consider BDTs made with a subset of the full set of variables.  
To be specific, we consider three different sets of variables:   
\begin{align}
	\label{eq:variable_sets}
	\begin{split}
		D \  \equiv \ &  \Bigl\{\log \theta_J, \nu_J,  \log \tau_1, \frac{\tau_2}{\tau_1}, 
					\frac{\tau_3}{\tau_2}, \frac{\tau_4}{\tau_3},   \\ &  \quad
								\bigl( \lambda_J,  \epsilon_J, \rho_J,  \delta_J \bigr)  \bigl|_{\text{C/A}},  		
			  				\bigl( \lambda_J,  \epsilon_J, \rho_J \bigr)  \bigl|_{k_T}  \Big\}  
	\end{split} \\
		D_{\text{C}} 	\  \equiv \  &  	\Big\{\log \theta_J, \nu_J \Big\}  \\
	\label{eq:variable_sets2}
          \begin{split}
		D_{\text{S}} 	\  \equiv \ &  \Bigl\{\log \tau_1, \frac{\tau_2}{\tau_1}, 
					\frac{\tau_3}{\tau_2}, \frac{\tau_4}{\tau_3},   \\ &  \quad
								\bigl( \lambda_J,  \epsilon_J, \rho_J,  \delta_J \bigr)  \bigl|_{\text{C/A}},  		
			  				\bigl( \lambda_J,  \epsilon_J, \rho_J \bigr)  \bigl|_{k_T}  \Big\}  		\; ,	
           \end{split}
\end{align}
where the subscripts $\CA$ or $\kt$ in Eqs.~\eqref{eq:variable_sets} and~\eqref{eq:variable_sets2}  imply that the observables are calculated using $\CA$ or $\kt$ subjets. The sets $D_{\text{C}} $ and $D_{\text{S}}$  consist of the conventional and the jet substructure variables respectively, whereas $D$ is the set of all variables. 

In a previous paper \cite{Ellis:Future} we described the more conventional separation of single photons from QCD-jets along with an initial introduction to the separation of single photons from photon-jets.  In both cases the single photons were treated as the signal.  Here we extend that discussion and focus on the photon-jets as the signal. We organize the results of our analysis into three subsections. First, we show the results of BDTs optimized to discriminate photon-jets from QCD-jets, the analogue of the seperation of single photons from QCD-jets. In the following subsection, we repeat the same study, but optimize it for treating  single photons as the background to photon-jets. Finally, we demonstrate how the BDTs might be used for an effective three-way separation of single photons from photon-jets from QCD-jets.

\subsection{\label{subsec:QCD-PJ} QCD-Jets as Background for Photon-jets}

\begin{figure}[h]
	\centering
	\includegraphics[width=0.49\textwidth]{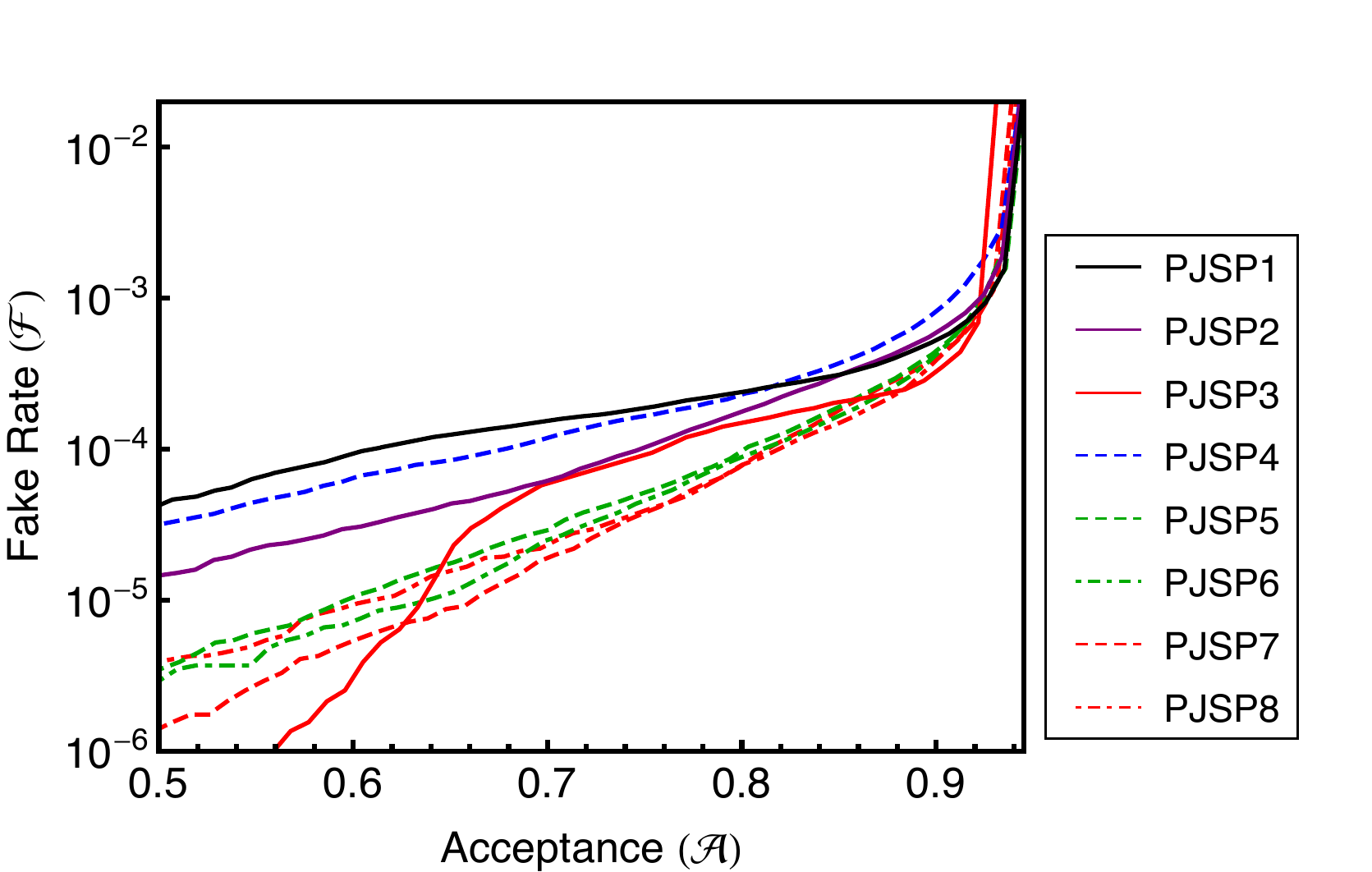}
	\caption{\label{fig:AF_PJvsQCD} The background fake rate versus signal acceptance where photon-jets from the different study points are the signal and QCD-jets are the background.  All variables in the set of discriminant $D$ are used in the analysis. }
\end{figure}

We use all of the variables in the set of discriminants $D$ in the BDTs in order to maximize the extraction of signal jets (photon-jets) from background (QCD-jets). This is similar to the separation of single photons from QCD-jets perfromed in Ref.~\cite{Ellis:Future}.   The BDTs are trained individually for each study point. The results for fake rate versus acceptance are shown in Fig.~\ref{fig:AF_PJvsQCD} for all of the study points.   In this plot the lower right is desirable and the upper left is undesirable.  Note that the acceptance rate for photon-jets is bounded above by about $0.94$ due to our preselection cut $\theta_J \geq 0.25$ (see Section~\ref{subsec:rho}).  The same cut eliminates approximately $98\%$ of the QCD-jets yielding a fake rate below $10^{-2}$ except at the largest acceptance.

For 2 photon photon-jets (study points 1 to 3) the separation becomes easier as $m_1$ increases yielding increasing separation between the photons inside the jet. The other physics scenarios tend to have even more structure within the photon-jets that the jet substructure variables allow us to use to suppress the QCD background. The more structure a jet possesses, the easier it becomes to discriminate it from (largely feature-less) QCD-jets. The conclusion from Fig.~\ref{fig:AF_PJvsQCD} is that, for photon-jets of varied kinematic features, we can achieve a very small QCD fake rate for a reasonably large acceptance rate. In more detail, for all of our study points a tagging efficiency (acceptance) of $\sim 70\%$ for photon-jets is accompanied by a fake rate for QCD-jets of only 1 in $10^4$ to 1 in $10^5$.

\begin{figure}[h]
	\centering
	\includegraphics[width=0.49\textwidth]{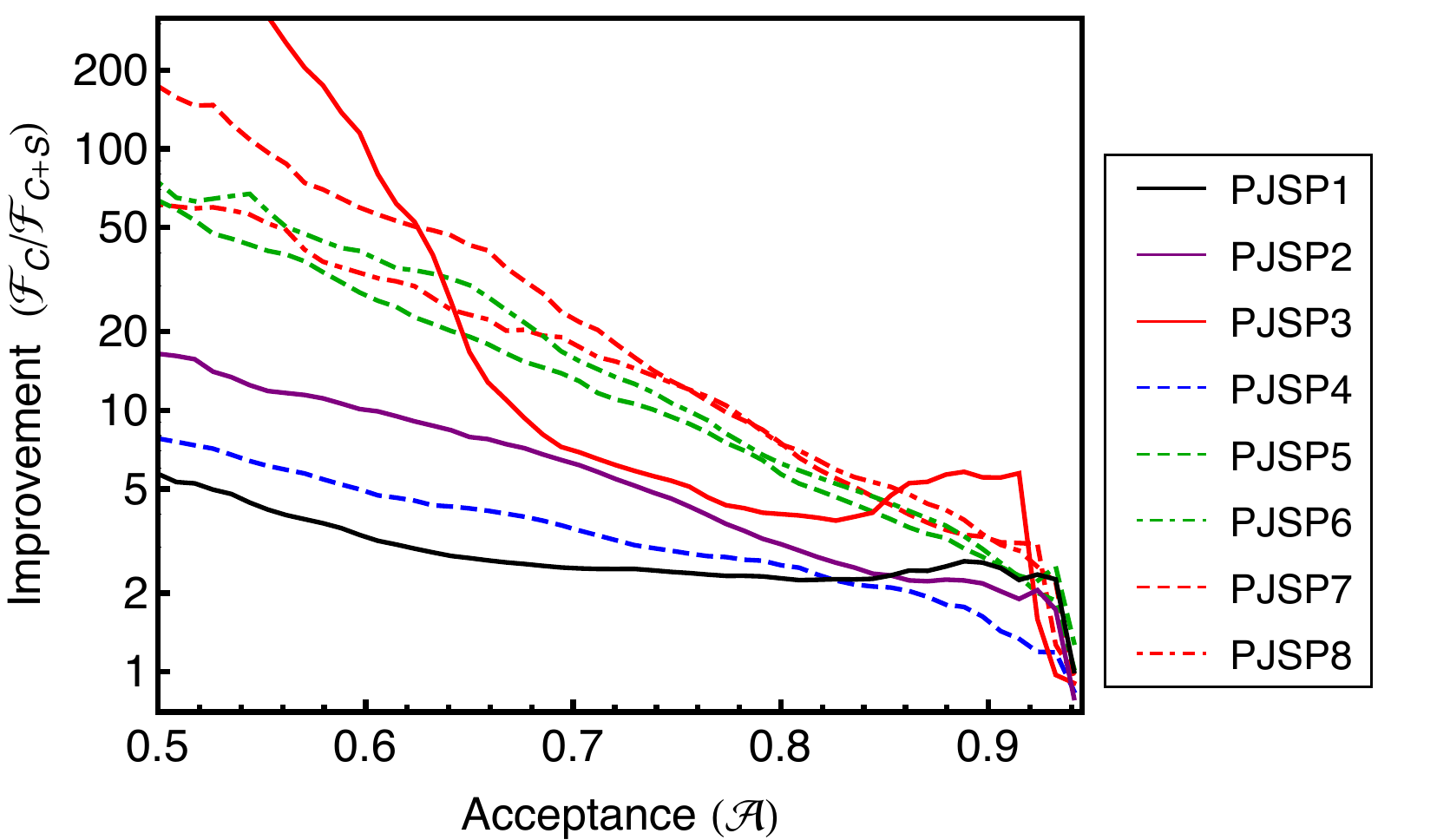}
	\caption{\label{fig:Imp_PJvsQCD} The improvement brought in because of the use of substructure variables are shown in the figure. For a quantitative definition of improvement see the text.}
\end{figure}

It is instructive to quantify the improvements made possible by including the jet substructure variables as discriminants.  To achieve this comparison we consider BDTs using only the conventional variables (i.e., we use the set $D_{\text{C}}$ of discriminants to train the BDTs). For a given acceptance of signal we thus obtain two different fake rates -- one when we use only the conventional variables (labeled $\mathcal{F}_\text{C}$), and another when we use conventional+jet substructure variables (labeled $\mathcal{F}_\text{C+S}$).  For a given acceptance, the ratio $\mathcal{F}_\text{C}/\mathcal{F}_\text{C+S}$ quantifies the improvement due to using jet substructure variables in this analysis. 
The improvement rates for conventional plus jet substructure variables over only conventional variables versus acceptance for discriminating photon-jets from QCD-jets is shown in Fig.~\ref{fig:Imp_PJvsQCD} for the different study points.  While Figs.~\ref{fig:hadfrac} and \ref{fig:chtrk} indicate that the conventional variables provide some discrimination between photon-jets and QCD-jets, Figs.~\ref{fig:tau} to \ref{fig:delta} indicate that the jet substructure variables provide a substantial number of new distinguishing features.
Fig.~\ref{fig:Imp_PJvsQCD} shows that these new features in the  jet substructure variables can provide substantial improvement.  Factors of 4 to 50 improvement in the discrimination of photon-jets from QCD-jets are possible at an acceptance of about $70\%$.    As expected more improvement is possible in those physics scenarios where the photon-jets have more structure.  Further, our results demonstrate that the use of jet substructure variables provides a tool to distinguish the different physics scenarios, i.e., the different study points, which is not possible with conventional variables alone.  

\subsection{\label{subsec:PvsPJ} Single Photons as Background to Photon-Jets}

\begin{figure}[h]
	\includegraphics[width=0.49\textwidth]{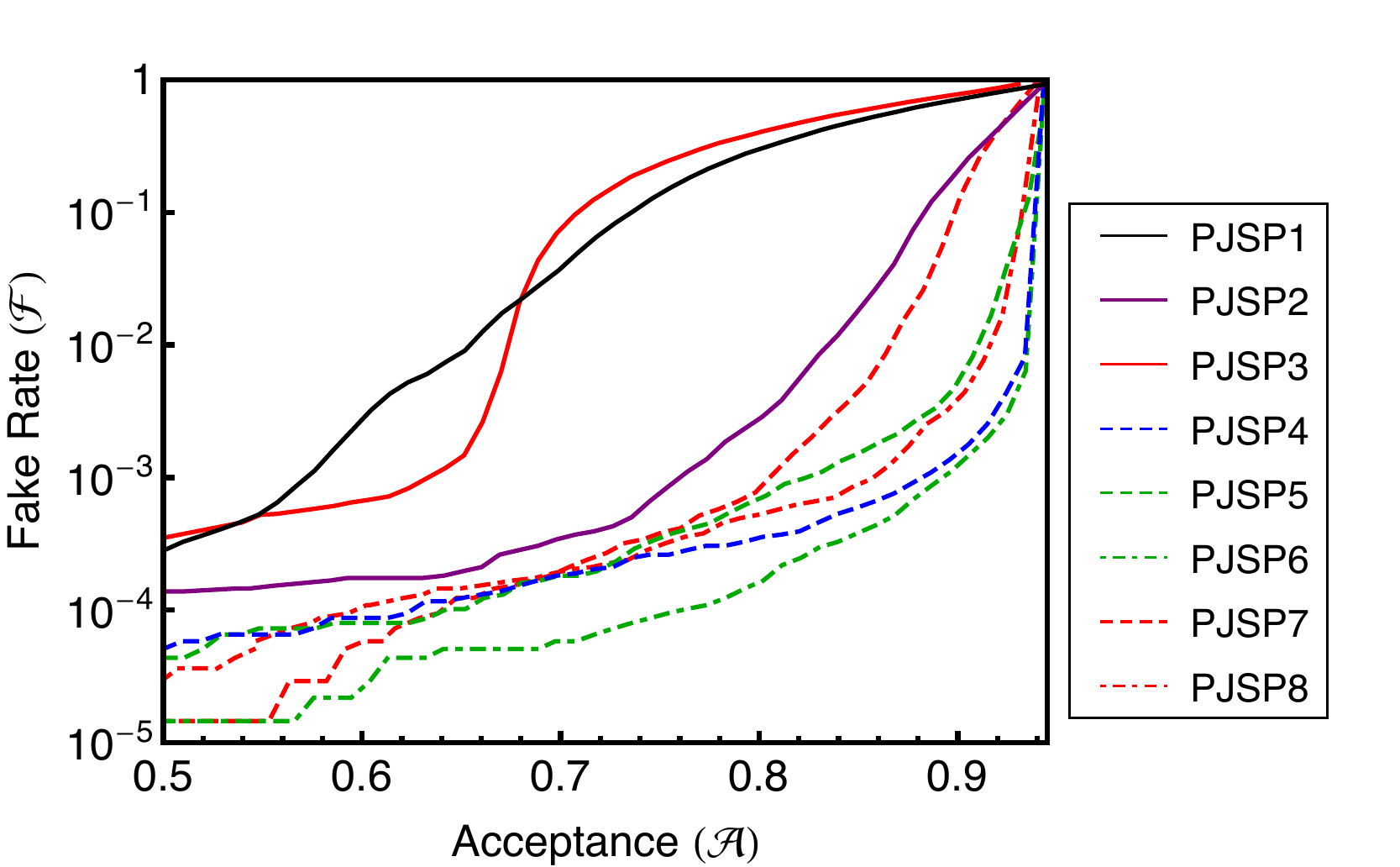}
	\caption{\label{fig:AF_PJvsP} The background fake rate versus signal acceptance curves are shown for all study points.  Here the photon-jets from the different study points are treated as the signal and single photons are the background. These curves employ all variables in the set of discriminants $D$.}
\end{figure}

Now consider the same analysis as in the previous section but with single photons treated as the background.  This new sort of separation is essential if we want to consider physics scenarios with photon-jets. Again we use all of the variables in the set of discriminants $D$ in the BDTs in order to maximize the extraction of signal jets (photon-jets) from background (single photons).  The BDTs are trained individually for each study point. The results for fake rate versus acceptance are shown in Fig.~\ref{fig:AF_PJvsP}.   As in  Fig.~\ref{fig:AF_PJvsQCD} the lower right is desirable and the upper left is undesirable.  Again the acceptance rate for photon-jets is bounded above by about $0.94$ due to our preselection cut $\theta_J > 0.25$ (see Section~\ref{subsec:rho}).  For the same reason a similar limit (0.94) holds also for the fake rate from single photons (although this is difficult to see on the logarithmic scale). 

The results in Fig.~\ref{fig:AF_PJvsP} teach us several lessons.  A photon-jet from \pjsp{1} consists of a pair of highly collinear photons. Such a jet is quite photon-like and thus difficult to separate from single photons.  Hence the corresponding (solid black) curve is most towards the upper left.  One needs to cut away almost half of the signal sample ($\mathcal{A} \sim 0.55$) in order to reduce the fake rate to 1 in $10^3$.    We also see that it is a challenge to separate the photon-jets for \pjsp{3} from single photons (the solid red curve).  In this scenario $m_1 = 10~\gev$ and the $n_1$ decays directly to two photons. Because of the large $m_1$ value, almost $30\%$ of these ($R=0,4$)  jets do not contain both of the photons from the $n_1$ decay, i.e., about $30\%$ of this jet sample are actually single photons (in the jet), and not photon-jets.  We saw this point earlier in essentially all of the individual jet substructure variable plots, Figs.~\ref{fig:tau} to \ref{fig:delta}, where the \pjsp{3} distribution exhibited an enhancement that overlapped with the corresponding peak in the single photon distribution. A larger separation of  \pjsp{3} from single photons can be obtained at an acceptance just below $0.7$, where these single photons configurations are cut away and the fake rate drops below 1 in $10^3$. The photon-jets of \pjsp{2} represent a `sweet' spot between \pjsp{1} and \pjsp{3} where the 2 photons are typically well enough separated to be resolved but close enough to be in the same jet.  Thus the \pjsp{2} (solid purple) curve is well below and to the right compared to the \pjsp{1} (solid black) and \pjsp{3} (solid red) curves.  Similarly the photon-jets at the other study points can be well separated at even larger acceptance rates using the combination of jet substructure and conventional discriminants.  For example, for the study points \pjsp{4} and  \pjsp{6}, even at $85\%$ acceptance, one obtains a fake rate \textit{smaller} than $1$ in $10^3$.   

Again it is instructive to determine the impact of the jet substructure variables for this analysis.  As in the previous subsection we consider BDTs using only the conventional variables (i.e., we use the set $D_{\text{C}}$ of discriminants to train the BDTs) to compare to the results from the full set $D$ of variables.  We plot the ratio of fake rates at fixed acceptance for these two analyses    
\begin{figure}[h]
	\includegraphics[width=0.49\textwidth]{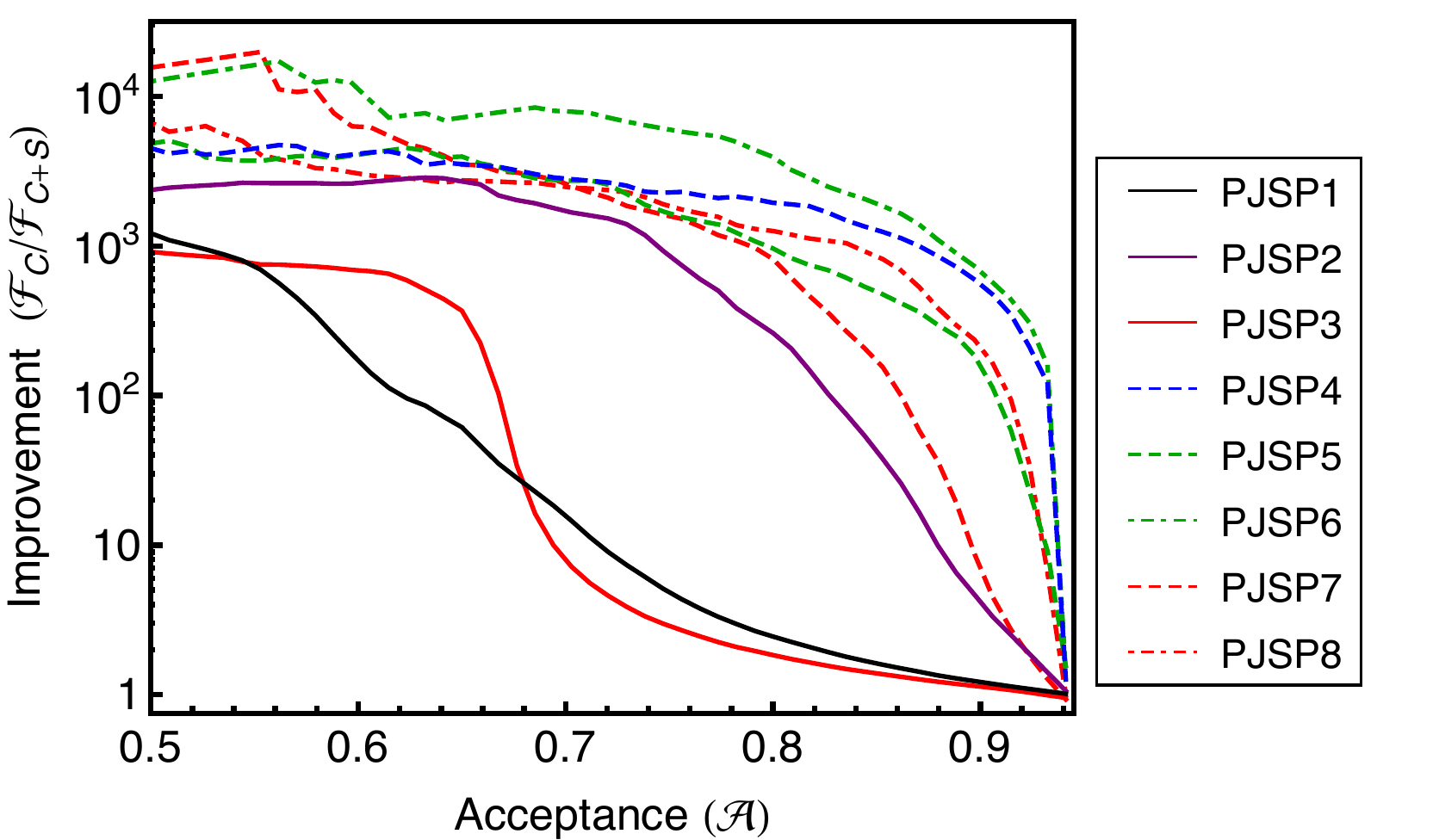}
	\caption{\label{fig:Imp_PJvsP}  The improvement brought in because of the use of substructure variables are shown in the figure. For a quantitative definition of improvement see the text.}
\end{figure}
in Fig.~\ref{fig:Imp_PJvsP} versus the acceptance. A comparison of Fig.~\ref{fig:Imp_PJvsP} and Fig.~\ref{fig:AF_PJvsP} indicates that the bulk of the separation of photon-jets from single photons is provided by the jet substructure variables, i.e., the improvement factor typically differs by less than a factor of 10 from one over the fake rate.  Further, the improvement factor ranges from 10 to more that $10^3$ even at acceptances as large as $90\%$ for all physics scenarios except \pjsp{1} and \pjsp{3}.  Even in these challenging cases substantial improvement is possible at lower acceptance rates.  This is not a surprise since the conventional variables are ineffective at distinguishing between photon-jets and single photons.  Recall from Fig.~\ref{fig:hadfrac} that the hadronic energy fraction distributions are nearly identical for photon-jets and single photons.  The distribution of the number of charge tracks associated with a jet, shown in Fig.~\ref{fig:chtrk}, also indicates only slight differences, arising from the somewhat different conversion rates for photon-jets versus single photons.  So it is clear that, if we want to be able to discriminate between photon-jets and single photons (and we do), the jet substructure variables provide the necessary tool.

\subsection{\label{subsec:QCDvsPJvsP} Three-way Separation}
Finally we come to the really interesting challenge: the \textit{simultaneous} separation all three samples: single photons, photon-jets, and QCD-jets.  In principle, one could perform three BDT training exercises, separating photon-jets from single photons, separating photon-jets from QCD-jets and separating single photons from QCD-jets, using one of the variable sets of Eqs.~\ref{eq:variable_sets} - \ref{eq:variable_sets2} in each case.  Then the responses from each of these BDTs for each jet could be used to separate the experimentally identified jets in the corresponding 3-dimensional `physics object' space.
In order to illustrate these ideas in a fairly simple analysis here we will focus on a two-dimensional analysis employing the two BDTs we have been discussing, separating photon-jets from single photons and separating photon-jets from QCD-jets.  There are still the related questions of which set of variables to use for each BDT and, in fact, how to characterize the `best separation'.\footnote{With three BDTs and the three BDT response numbers for each jet, the `best separation' presumably corresponds to the three distinct physics objects being sent to three diagonally opposite vertices of the BDT response cube (on a equilateral triangle with side of length $\sqrt{2}$ times the length of the edge of the cube).}  Qualitatively at least, we find good 2-dimensional separation for the following definitions of the BDTs. 
One is trained to separate QCD-jets and photon-jets based only on the conventional discriminants ($D_\text{C}$) and is plotted on the vertical axis in the following plots, while the other BDT is trained to separate photon-jets from single photons with the substructure discriminants ($D_\text{S}$) alone and is plotted along the horizontal axis.  We present the results in terms of two-dimensional contour plots where the numerical values associated with  a given contour corresponds to the relative probability to find a calorimeter object of the given kind (indicated by the color) in a cell of size $0.1 \times 0.1$ in BDT response units.   (Note that, by construction, the BDT responses have values in the range $-1$ to $+1$, where $+1$ means `signal-like' and $-1$ means `background-like'.)  The color coding in these figures matches the previous choices.  Red is for QCD-jets, blue for photon-jets and green is for single photons. 

\begin{figure}[h]
\includegraphics[width=0.45\textwidth]{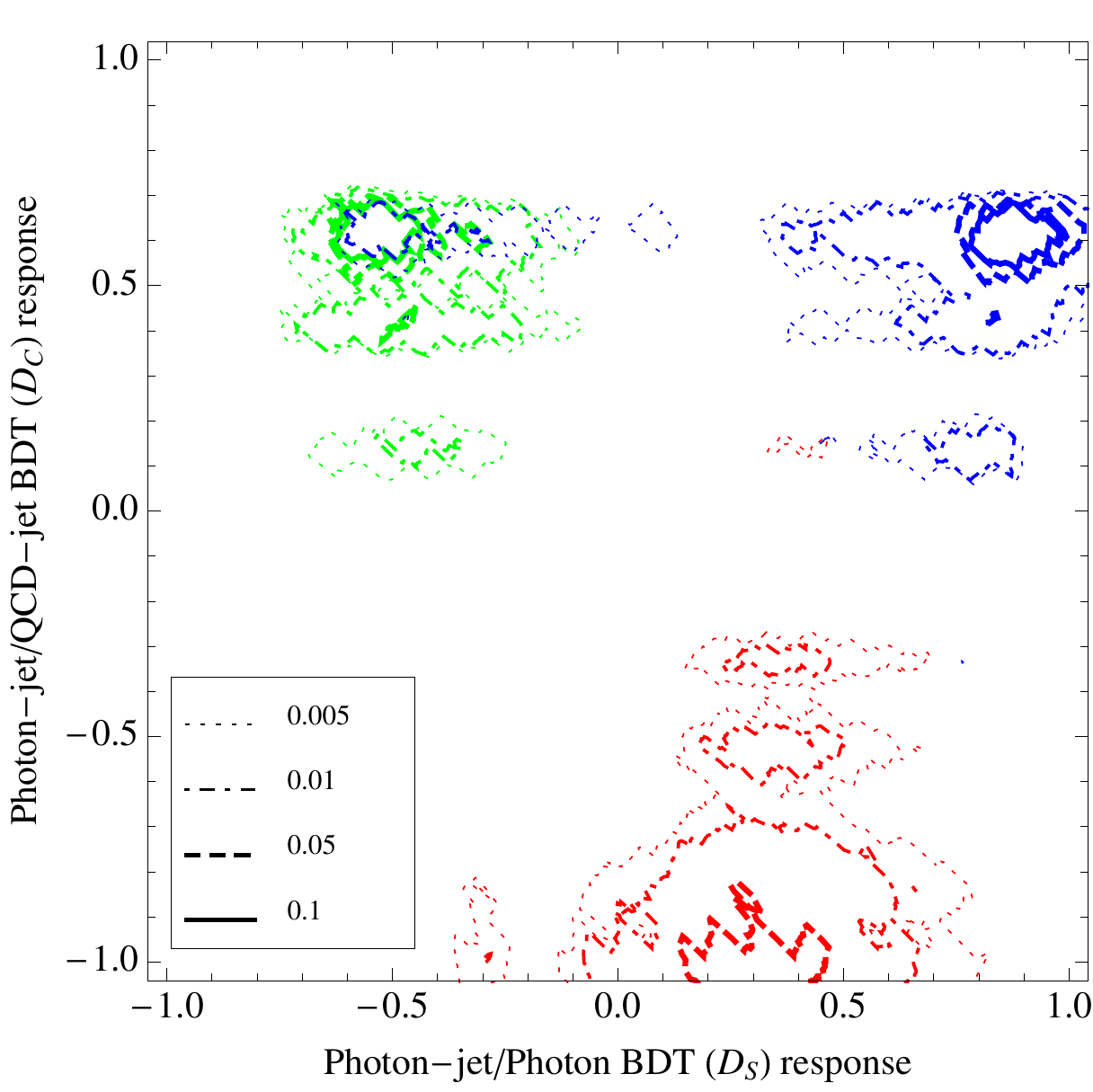} 
\caption{\label{fig:Density_2JPJ} The BDT responses of QCD-jets(red), single photons(green) and photon-jets(blue) for photon-jets at \pjsp{2}. The $D_S$ variables are used on the horizontal axis and the $D_C$ variables on the vertical axis.}
\end{figure}

As a first example,  Fig.~\ref{fig:Density_2JPJ} indicates the 2-dimensional distributions resulting from the BDTs for \pjsp{2}, a scenario with typically two photons in the photon-jet with small angular separation due to the small value of $m_1$.  When interpreting the following figures it is important to recall that the jet samples indicated in these figures are constrained to satisfy $\theta_J \leq 0.25$, which means that we are only keeping the approximately $2\%$ of QCD-jets that are most `photon-like'.  
However,  Fig.~\ref{fig:Density_2JPJ} indicates a pretty clear separation between the QCD-jets and the true photon objects (little red above $0.0$ in the vertical direction).  On the other hand, as we expect from our previous one-dimensional discussions in Subsection~\ref{subsec:PvsPJ}, the blue (photon-jet) contours in the upper-left green (single photon) region indicate that it is a challenge to completely separate (\pjsp{2}) photon-jets from single photons. 
\begin{figure}[h]
\includegraphics[width=0.45\textwidth]{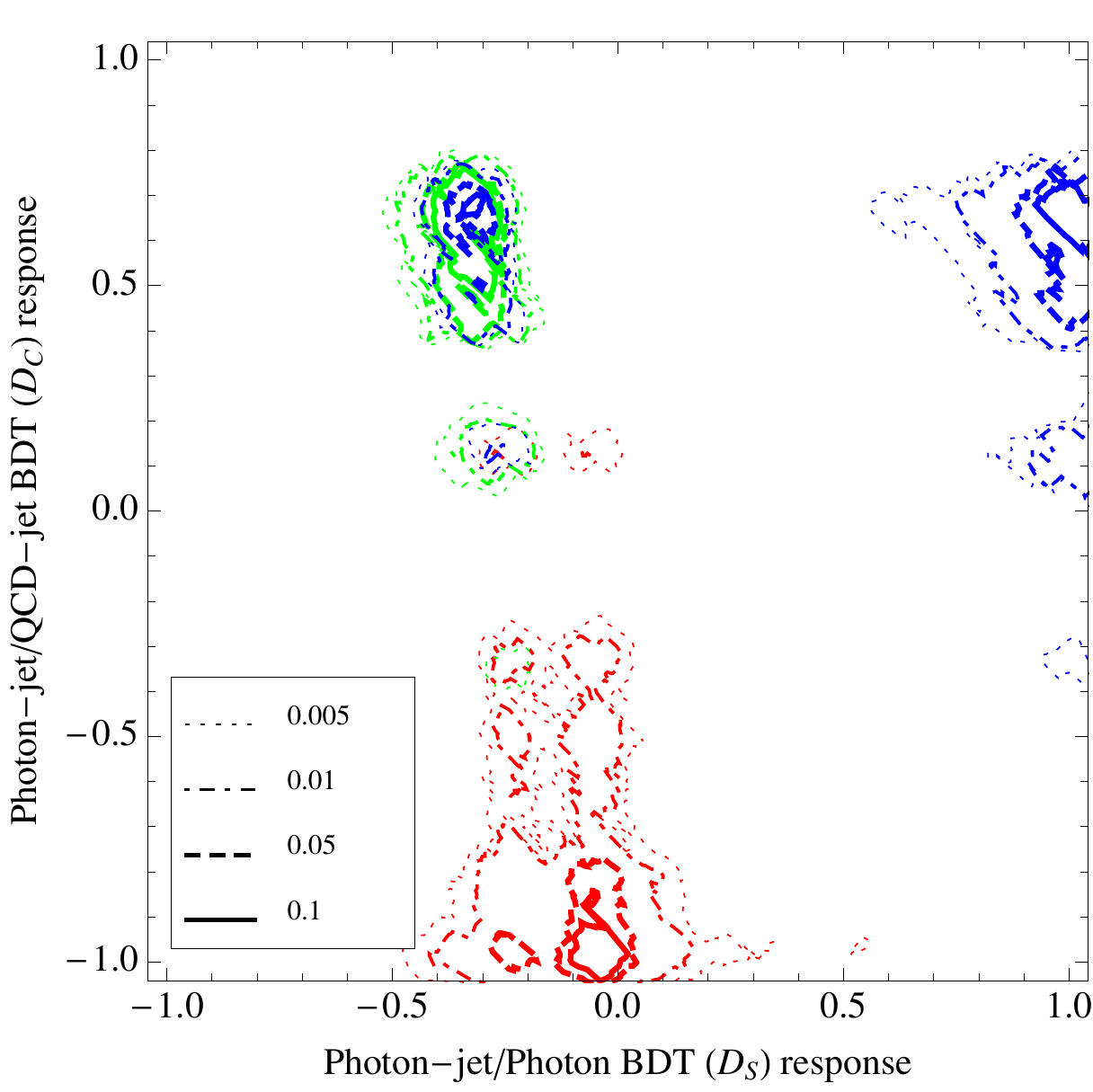}
	\caption{\label{fig:Density_3JPJ} The BDT responses of QCD-jets(red), single photons(green) and photon-jets(blue) for photon-jets at \pjsp{3}.}
\end{figure}

In the case of \pjsp{3} photon-jets, as indicated in Fig.~\ref{fig:Density_3JPJ}, the photon-jet versus single photon separation challenge is even larger, as we have already discussed. Again we have photon-jets with potentially two photons but, due to the relatively large $m_1$ value, one of those photons is sometimes outside of the identified jet.  This explains the small region with a solid blue (probability $~0.1$) contour inside the green (single photon) region. 

\begin{figure}[h]
\includegraphics[width=0.45\textwidth]{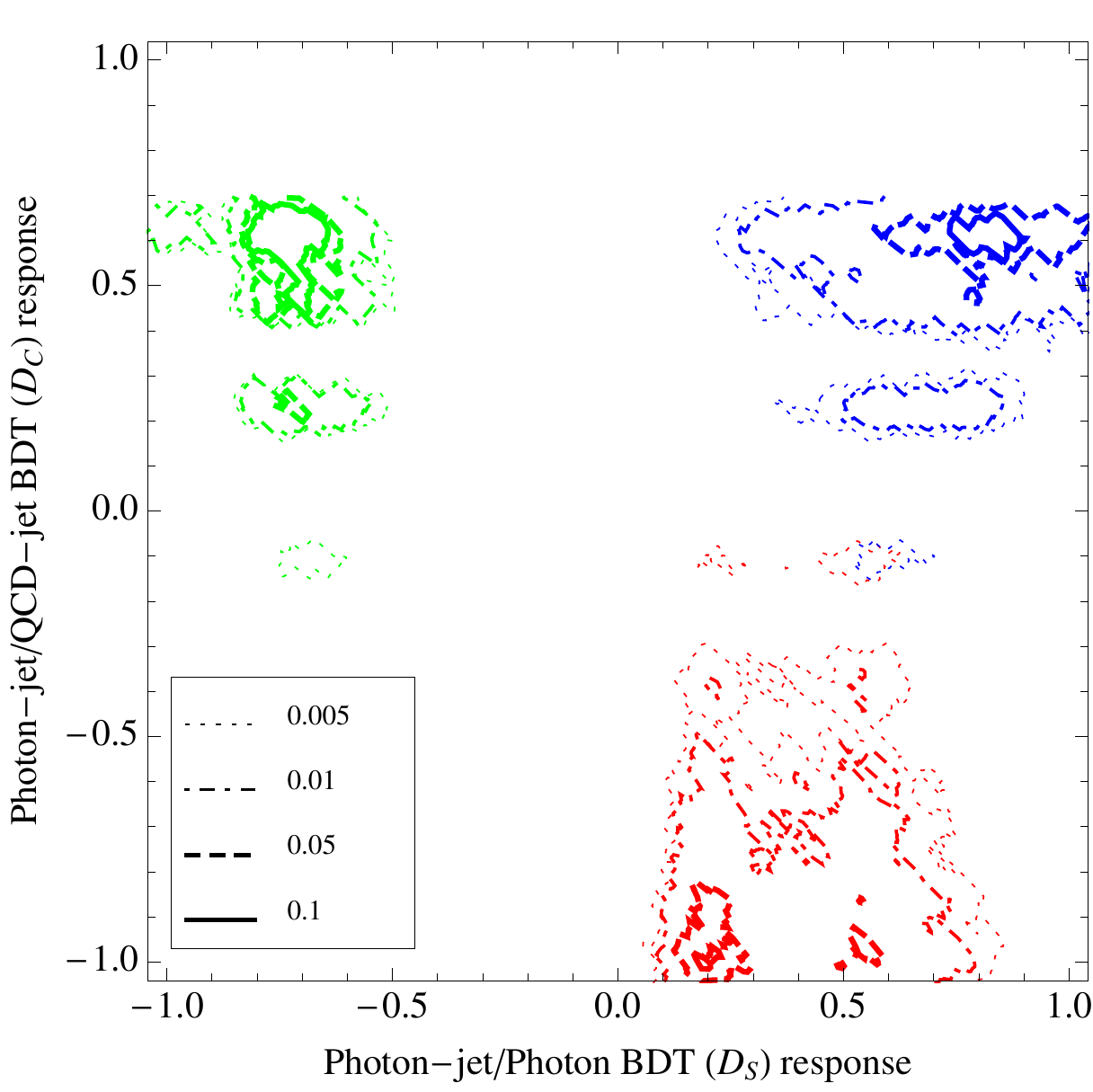} 
\caption{\label{fig:Density_4JPJ} The BDT responses of QCD-jets(red), single photons(green) and photon-jets(blue) for photon-jets at \pjsp{4}.}
\end{figure}
\begin{figure}[h]
\includegraphics[width=0.45\textwidth]{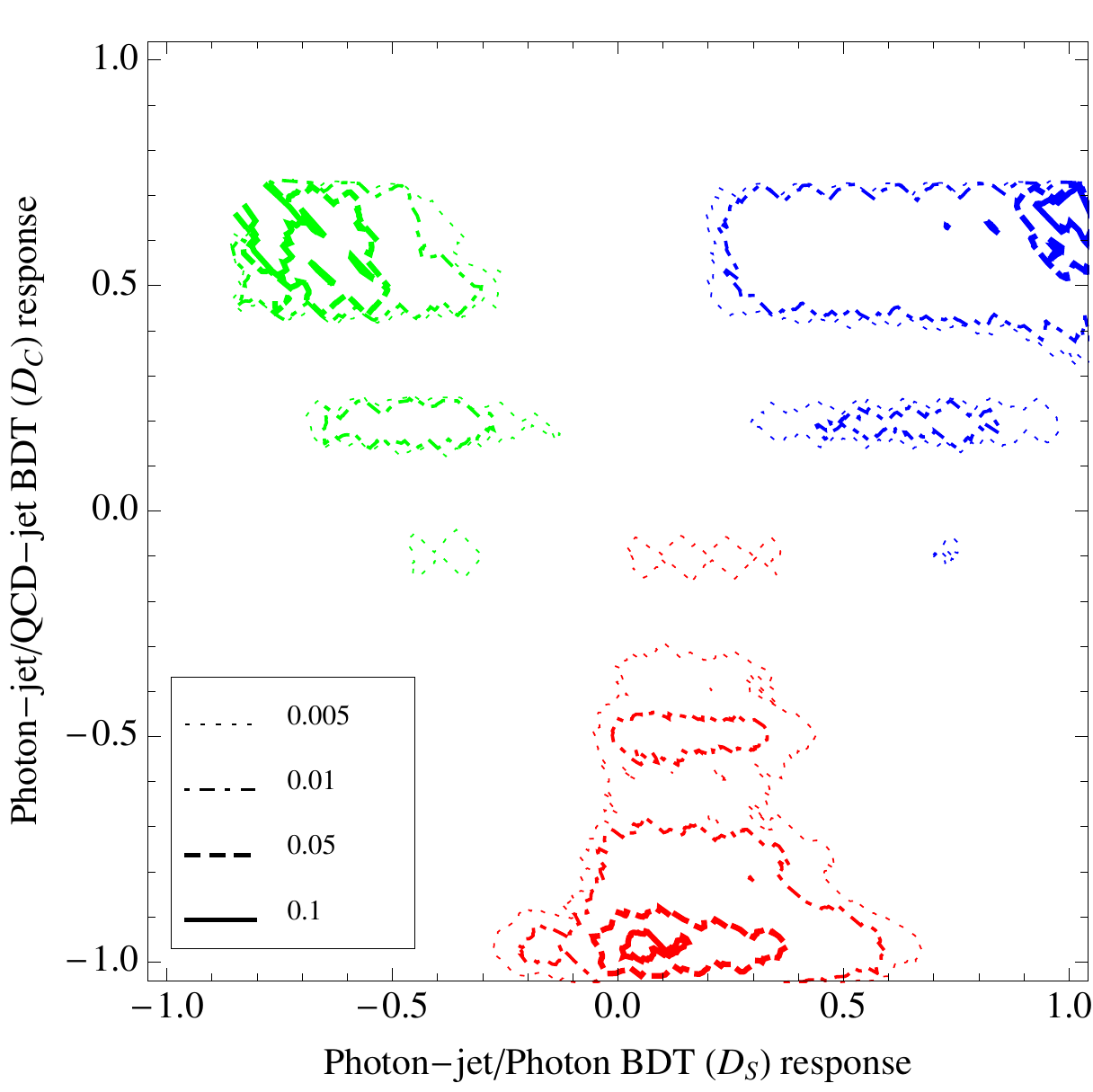} 
	\caption{\label{fig:Density_8JPJ}  The separation of QCD-jets (red), single photons (green) and photon-jets(blue) for photon-jets at \pjsp{8}.}
\end{figure}
The corresponding results for the more complex (and more easily separated) photon-jets of \pjsp{4} and \pjsp{8}, typically with 4 photons in a photon-jet, are displayed in Figs.~\ref{fig:Density_4JPJ} and  \ref{fig:Density_8JPJ}.  In these scenarios the three-way photon-jet versus single photon versus QCD-jet separation is fairly cleanly achieved using just the $D_S$ (horizontal) and $D_C$ (vertical) variable sets.   At the $0.005$ level there is only a tiny overlap of photon-jets with QCD-jets for \pjsp{4} (near the location ($0.5$,$0.0$) in Fig.~\ref{fig:Density_4JPJ}) and no overlap for \pjsp{8} (Fig.~\ref{fig:Density_8JPJ}).

\begin{figure}[h]
\includegraphics[width=0.45\textwidth]{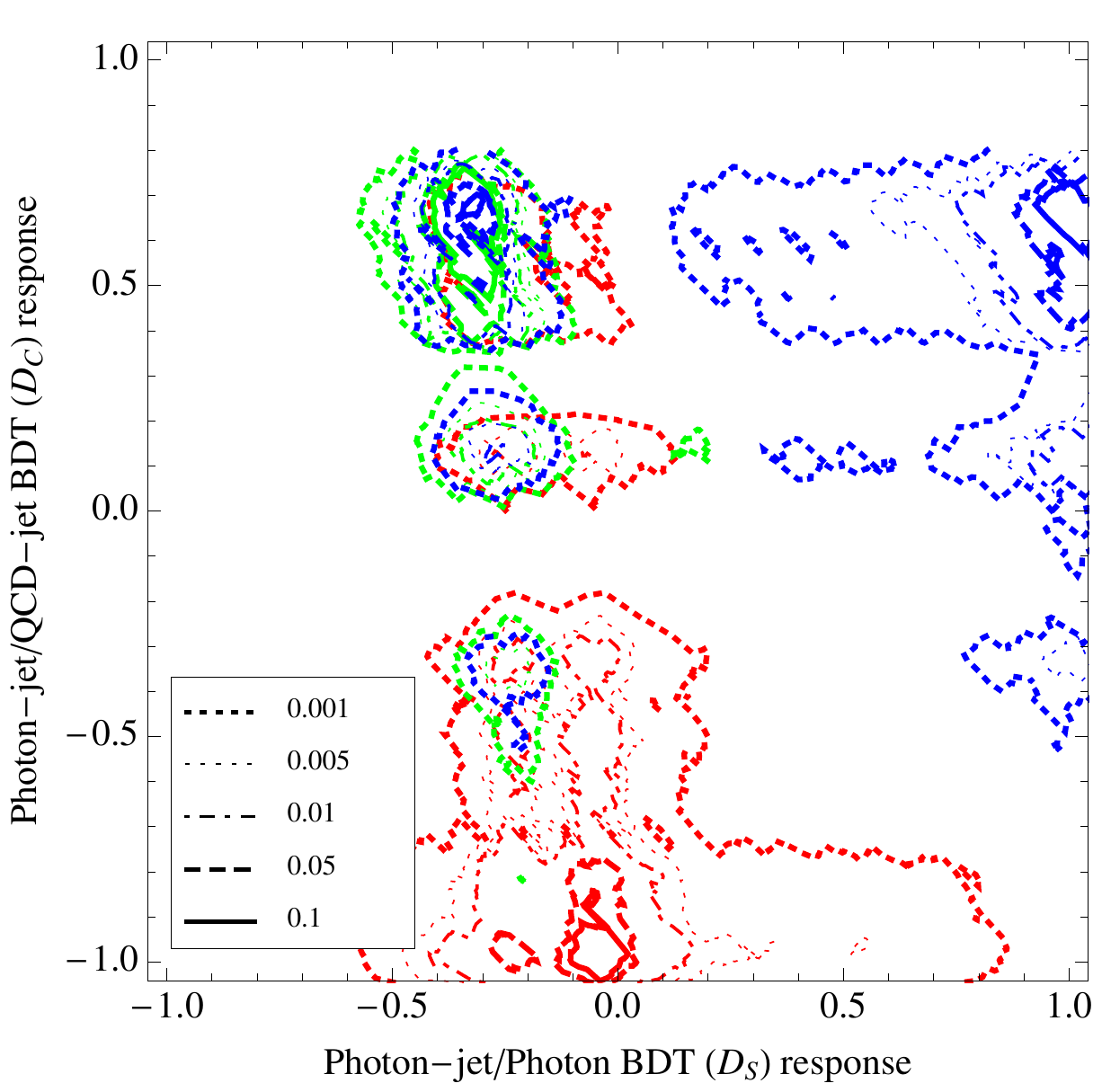}
	\caption{\label{fig:Density_3JPJ_ex} The BDT responses of QCD-jets(red), single photons(green) and photon-jets(blue) for photon-jets at \pjsp{3} including a contour at $0.001$.}
\end{figure}
\begin{figure}[h]
\includegraphics[width=0.45\textwidth]{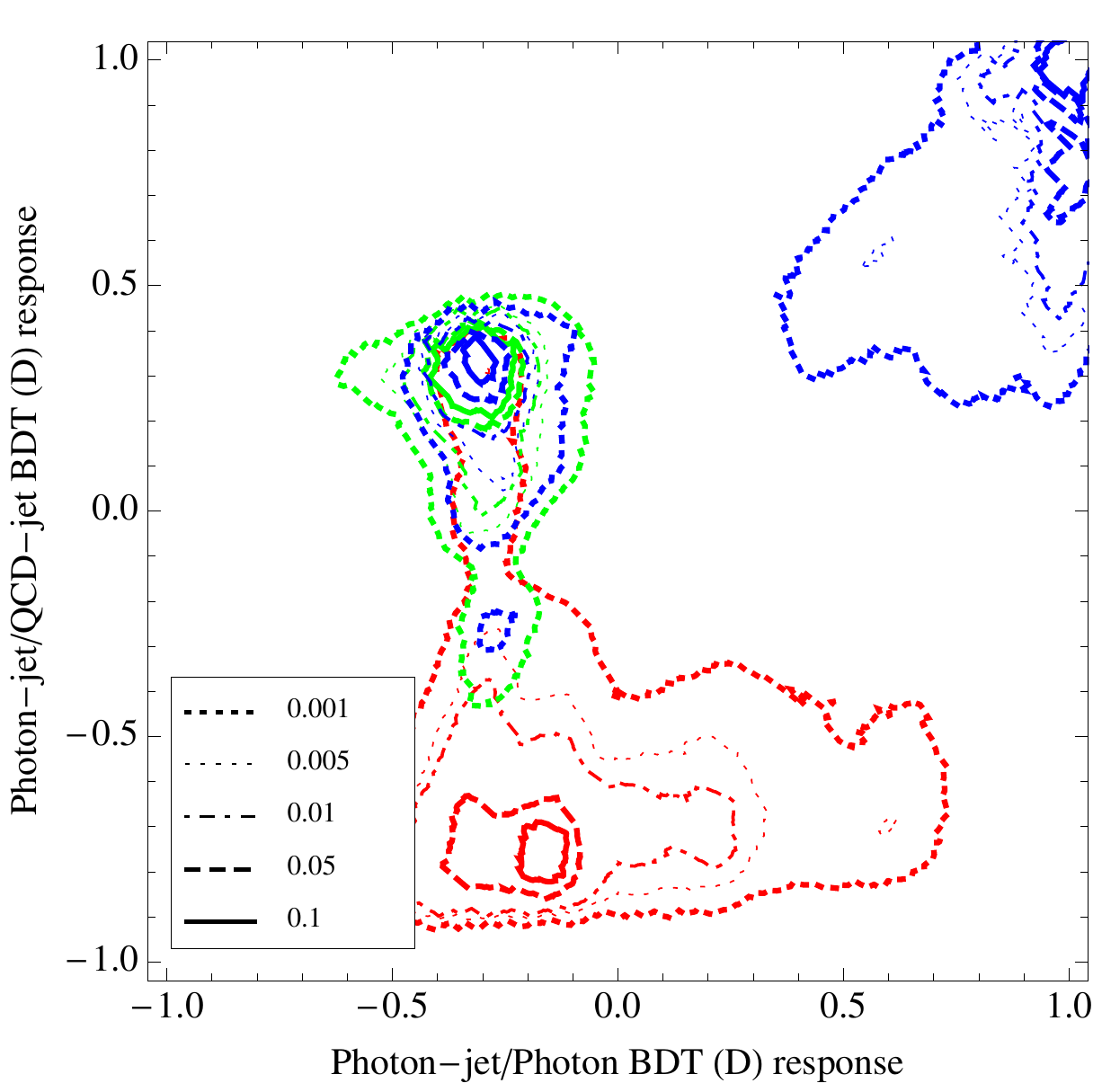}
	\caption{\label{fig:Density_3JPJ_D} The BDT responses of QCD-jets(red), single photons(green) and photon-jets(blue) for photon-jets at \pjsp{3} using the full set $D$ of variables on the horizontal axis.}
\end{figure}

Before ending this section we should discuss one other point.  From our previous discussion, one would expect to improve the photon-jet versus single photon separation by using the full $D$ set of variables (instead of the $D_S$ variables alone), and this expectation raises one of the interesting, and challenging, features of \textit{simultaneous} separations.  Since we are currently training the BDTs so that each BDT separates one type of signal from one type of background, while, at the same time, trying to perform a three-way separation, it can happen that an improvement in one separation corresponds to a degradation in another of the separations.  To illustrate this point we first reproduce the results in Fig.~\ref{fig:Density_3JPJ}, but now include a contour at relative probability $0.001$, which we did not include earlier to avoid plots that are too busy.  The resulting plot is shown in Fig.~\ref{fig:Density_3JPJ_ex}.  Now we perform the same analysis but using the full variable set $D$ in both BDTs.  The resulting contour plot is displayed in Fig.~\ref{fig:Density_3JPJ_D}, which illustrates the relevant points.  The $0.001$ level boundaries for single photons (green) and photon-jets (blue) are now somewhat better separated, although the effectively one-photon-jets from \pjsp{3} (when one of the photons is very soft or is outside of the jet) still lie within the single photon boundary.
At the same time, however, the separation between single photons (green) and the (typically more numerous) QCD-jets (red) is somewhat degraded 
(the green and red regions have moved towards each other).  Due to the coupling between the different pairwise separations, optimizing such a three-way separation takes careful work and likely depends on the details of the actual analysis and detector. 

These results clearly suggest that a three-way separation is possible, including the ability to distinguish different photon-jet scenarios.  Further enhancement will arise from using the full 3-dimensional structure and from using a realistic detector simulation in the training.  A thorough optimization in the context of a real detector and actual data may select different, more effective choices of the discriminating variables.

\section{\label{sec:conclusion} Conclusion}

In this paper we have attempted to link several concepts, some conventional and some less so, with the goal of enhancing the searches for and the analyses of both Standard Model and Beyond the Standard Model physics.  We advocate employing general techniques for analyzing and interpreting the detector objects identified by applying standard jet algorithms to the calorimeter cells of typical hadron collider detectors, allowing a universal language for such objects.  We have demonstrated the efficacy of employing the recent developments in jet substructure techniques to separate and identify these detector objects in terms of physics objects.  Continuing the efforts begun in Ref.~\cite{Ellis:Future}, we have focused on identifying three specific physics objects, the familiar single photons, QCD-jets and the Beyond the Standard Model (and LHC) relevant photon-jets.  In particular, we have demonstrated that it is possible to achieve significant separation between photon-jets and their dominant backgrounds, i.e., single photons and QCD-jets. We expect that both the ATLAS and CMS groups could enhance their searches for signatures of new physics by adopting the methods described.   These methods should allow the separation of photon-jets from single photons from QCD-jets, and also provide some identification of the specific dynamics yielding the photon-jets.

We note that our simulation does not take into account the impact of magnetic fields inside the detectors.  On the other hand, one might interpret this absence of  a magnetic field as making our results more conservative. When the magnetic field bends the electrons and positrons from converted photons, this serves to generate more structure inside the jet. The substructure variables, as we have described, tend to become more powerful with more structure. A more detailed analysis is, however, beyond the scope of this paper.  

Finally, it is worth mentioning that the formalism and techniques developed in this  paper for photon-jets should work in a similar way for the case of collinear electrons, often labeled `electron-jets'~\cite{Ruderman:2009tj, Cheung:2009su, Falkowski:2010cm, Falkowski:2010gv}. An electron-jet is characterized by a large number of charged tracks along with a small hadronic energy fraction.  Also, we expect the electrons inside these jets to bend in a magnetic field,
creating more substructure.  Therefore we anticipate that multivariate analyses similar to those described here will be correspondingly effective at separating electron-jets from QCD-jets (and photon-jets).

\section*{Acknowledgements}
The authors would like to acknowledge stimulating conversations with  Henry Lubatti and Gordon Watts regarding the project. We especially thank Henry Lubatti for his careful reading of the manuscript.  TSR would like to thank the hospitality of CERN, where part of the work was completed. The work of SDE, TSR and JS was supported, in part, by the US Department of Energy under contract numbers DE-FGO2-96ER40956. JS would also like to acknowledge partial support from a DOE High Energy Physics Graduate Theory Fellowship.  Computing resources were provided by the University of Washington supported by the US National Science Foundation contract ARRA-NSF-0959141.

\appendix


\section{Technical Details}

\subsection{\label{sec:app-conversion} Conversions}
\begin{figure}[h]
\includegraphics[scale=0.5]{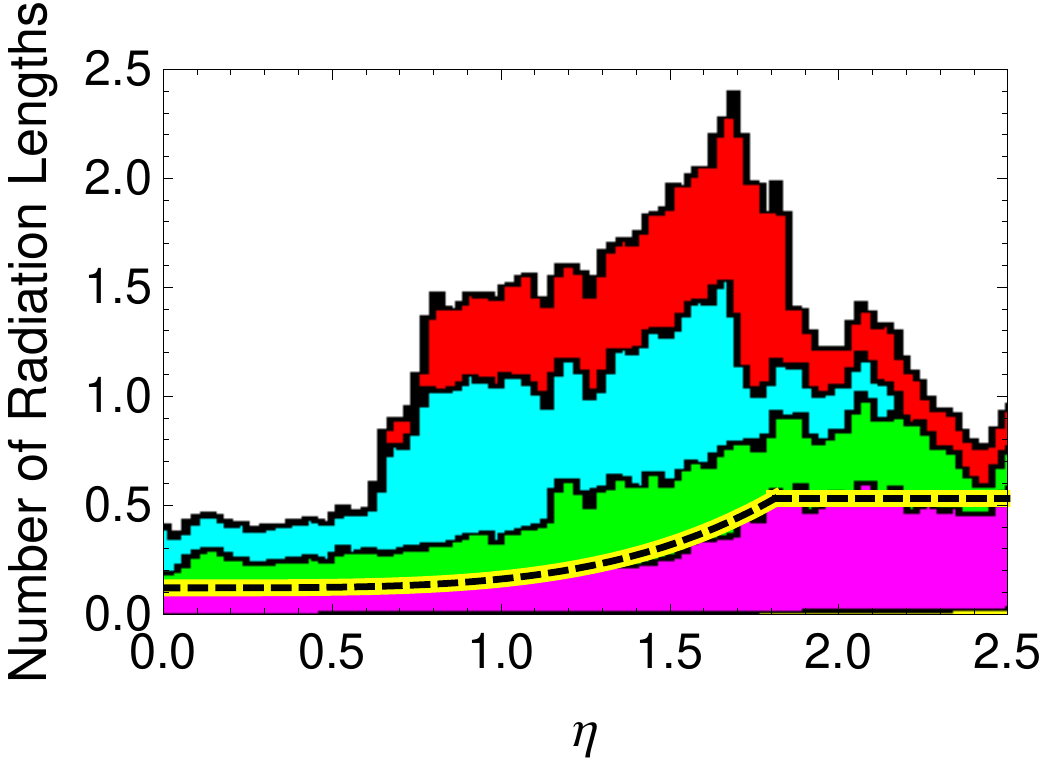}
\caption{\label{fig:conversion} Number of radiation length up to and through the layer of pixel detectors in the ATLAS detector. }
\end{figure}

In the material of the detector photons convert into electron-positron pairs.  Our implementation of photon conversion is based on the properties of the ATLAS detector\cite{Aad:2008zzm}. We associate an $\eta$ dependent probability of conversion with every photon.  This probability is a function of the number of radiation lengths a photon passes through (labeled $n(\eta)$) in order to escape the first layer of the pixels. We model the probability to convert using the expression:
\begin{equation}
P(\eta) = 1 - \exp \left(-\frac{7}{9}n(\eta)\right) \, ,
\end{equation} 
where  the factor of $7/9$ comes from conversion between radiation length and mean free path.  A plot of the extracted radiation length profile of the inner pixel detector in the ATLAS detector that we use to determine $n(\eta)$ is displayed in Fig.~\ref{fig:conversion}.
The yellow-black dashed line shows the extracted $\eta$-dependent number-of-radiation-lengths a photon needs to travel before it exits the pixel layer.  If the photon converts before the dashed line, it is treated as a charged track, otherwise no charged track is included in our simulation.

\subsection{\label{sec:Calorimeter} Details of the Calorimeters} 
Our simulation of calorimeters closely resembles the calorimeters implemented in the widely used simulation tool PGS~\cite{PGS4}.  The electromagnetic calorimeter ECal covers $ \left| \eta \right| \leq 2.5$ in the pseudorapidity direction, whereas the hadronic calorimeter HCal covers the range $ \left| \eta \right| \leq 5.0$. Both of the calorimeters provide a full $2\pi$ coverage in $\phi$, the azimuthal direction. The ECal has granularity of $0.025\times 0.025$ in the $\eta$-$\phi$ plane. The HCal, on the other hand, has a coarser granularity with cells of size $0.1\times 0.1$.

We assume that all particles generated at the interaction point (i.e., the Pythia output), except charged particles with energy less than $0.1~\gev$, reach the ECal.  Photon/electrons within $ \left| \eta \right| \leq 2.5$ deposit $99\%$ of their energy in the ECal and the rest in the HCal.  The ECal also fully absorbs hadrons with energy less than $0.5~\gev$. For more energetic hadrons,  $0.5~\gev$ is deposited in the ECal and the rest in the HCal. Photon/electron/hadrons that lie outside the pseudorapidity range of the ECal but within the range of the HCal (i.e., $5.0 \geq \left| \eta \right|  > 2.5$)  are completely absorbed in the HCal.  

In our simulation, muons also deposit energy in the calorimeters.  Within the range $ \left| \eta \right| \leq 2.5$, muons with   $E < 0.5~\gev$ deposit all of their energy in the ECal; muons with $0.5~\gev < E < 2.5~\gev$ deposit $0.5~\gev$ in the ECal and the rest in the HCal;  muons with $E > 2.5~\gev$ deposit $0.5~\gev$ in the ECal  and  $2.0~\gev$ in the HCal. For muons within the rapidity range $5.0 \geq \left| \eta \right|  > 2.5$, those with $E < 2.0~\gev$ are fully absorbed in the HCal, while muons with $E > 2.0~\gev$ deposit $2.0~\gev$ in the HCal.

\subsection{\label{sec:app-Moliere} Moliere Matrix}

In order to simulate the transverse smearing  in the ECal, we deposit the energy of each particle not just into the specific cell through which it passes, but also into the surrounding cells. The  ($\eta$-$\phi$)  coordinates of a particle determine the cell in which it deposits most of its energy (call it the ($i,j$)-th cell in the grid). We distribute its energy in the neighboring cells  according to the following table:
\begin{center}
\begin{tabular}{c|c|c|c}
     & i-1   &    i  & i+1   \\ \hline
 j-1 & 0.00(4) & 0.01(4) & 0.00(4) \\ \hline
 j   & 0.01(4) & 0.92(4) & 0.01(4) \\ \hline
 j+1 & 0.00(4) & 0.01(4) & 0.00(4)   
\end{tabular}
\end{center}
We estimate these numbers by integrating the energy deposited in an electromagnetic shower in a cell of size ($0.025\times 0.025$) and situated $1.5~\text{m}$ from the origin of the shower.  We assume that the cell is made of lead ($R_M(\text{Pb}) = 1.6~\text{cm}$) .


\begin{thebibliography}{60}
\expandafter\ifx\csname natexlab\endcsname\relax\def\natexlab#1{#1}\fi
\expandafter\ifx\csname bibnamefont\endcsname\relax
  \def\bibnamefont#1{#1}\fi
\expandafter\ifx\csname bibfnamefont\endcsname\relax
  \def\bibfnamefont#1{#1}\fi
\expandafter\ifx\csname citenamefont\endcsname\relax
  \def\citenamefont#1{#1}\fi
\expandafter\ifx\csname url\endcsname\relax
  \def\url#1{\texttt{#1}}\fi
\expandafter\ifx\csname urlprefix\endcsname\relax\def\urlprefix{URL }\fi
\providecommand{\bibinfo}[2]{#2}
\providecommand{\eprint}[2][]{\url{#2}}

\bibitem[{\citenamefont{Aad et~al.}(2012)}]{:2012gk}
\bibinfo{author}{\bibfnamefont{G.}~\bibnamefont{Aad}} \bibnamefont{et~al.}
  (\bibinfo{collaboration}{ATLAS Collaboration}),
  \bibinfo{journal}{Phys.Lett.B}  (\bibinfo{year}{2012}), \eprint{1207.7214}.

\bibitem[{\citenamefont{Chatrchyan et~al.}(2012)}]{:2012gu}
\bibinfo{author}{\bibfnamefont{S.}~\bibnamefont{Chatrchyan}}
  \bibnamefont{et~al.} (\bibinfo{collaboration}{CMS Collaboration}),
  \bibinfo{journal}{Phys.Lett.B}  (\bibinfo{year}{2012}), \eprint{1207.7235}.

\bibitem[{\citenamefont{Dobrescu et~al.}(2001)\citenamefont{Dobrescu,
  Landsberg, and Matchev}}]{Dobrescu:2000jt}
\bibinfo{author}{\bibfnamefont{B.~A.} \bibnamefont{Dobrescu}},
  \bibinfo{author}{\bibfnamefont{G.~L.} \bibnamefont{Landsberg}},
  \bibnamefont{and} \bibinfo{author}{\bibfnamefont{K.~T.}
  \bibnamefont{Matchev}}, \bibinfo{journal}{Phys.Rev.}
  \textbf{\bibinfo{volume}{D63}}, \bibinfo{pages}{075003}
  (\bibinfo{year}{2001}), \eprint{hep-ph/0005308}.

\bibitem[{\citenamefont{Toro and Yavin}(2012)}]{Toro:2012sv}
\bibinfo{author}{\bibfnamefont{N.}~\bibnamefont{Toro}} \bibnamefont{and}
  \bibinfo{author}{\bibfnamefont{I.}~\bibnamefont{Yavin}}
  (\bibinfo{year}{2012}), \eprint{1202.6377}.

\bibitem[{\citenamefont{Draper and McKeen}(2012)}]{Draper:2012xt}
\bibinfo{author}{\bibfnamefont{P.}~\bibnamefont{Draper}} \bibnamefont{and}
  \bibinfo{author}{\bibfnamefont{D.}~\bibnamefont{McKeen}},
  \bibinfo{journal}{Phys.Rev.} \textbf{\bibinfo{volume}{D85}},
  \bibinfo{pages}{115023} (\bibinfo{year}{2012}), \eprint{1204.1061}.

\bibitem[{ATL(2012{\natexlab{a}})}]{ATLAS-CONF-2012-079}
\bibinfo{type}{Tech. Rep.} \bibinfo{number}{ATLAS-CONF-2012-079},
  \bibinfo{institution}{CERN}, \bibinfo{address}{Geneva}
  (\bibinfo{year}{2012}{\natexlab{a}}).

\bibitem[{\citenamefont{Seymour}(1994)}]{Seymour:1993mx}
\bibinfo{author}{\bibfnamefont{M.~H.} \bibnamefont{Seymour}},
  \bibinfo{journal}{Z.Phys.} \textbf{\bibinfo{volume}{C62}},
  \bibinfo{pages}{127} (\bibinfo{year}{1994}).

\bibitem[{\citenamefont{Brooijmans}(2008)}]{Brooijmans:1077731}
\bibinfo{author}{\bibfnamefont{G.}~\bibnamefont{Brooijmans}},
  \bibinfo{type}{Tech. Rep.} \bibinfo{number}{ATL-PHYS-CONF-2008-008.
  ATL-COM-PHYS-2008-001}, \bibinfo{institution}{CERN},
  \bibinfo{address}{Geneva} (\bibinfo{year}{2008}).

\bibitem[{\citenamefont{Butterworth et~al.}(2007)\citenamefont{Butterworth,
  Ellis, and Raklev}}]{Butterworth:2007ke}
\bibinfo{author}{\bibfnamefont{J.}~\bibnamefont{Butterworth}},
  \bibinfo{author}{\bibfnamefont{J.~R.} \bibnamefont{Ellis}}, \bibnamefont{and}
  \bibinfo{author}{\bibfnamefont{A.}~\bibnamefont{Raklev}},
  \bibinfo{journal}{JHEP} \textbf{\bibinfo{volume}{0705}}, \bibinfo{pages}{033}
  (\bibinfo{year}{2007}), \eprint{hep-ph/0702150}.

\bibitem[{\citenamefont{Butterworth
  et~al.}(2008{\natexlab{a}})\citenamefont{Butterworth, Davison, Rubin, and
  Salam}}]{Butterworth:2008iy}
\bibinfo{author}{\bibfnamefont{J.~M.} \bibnamefont{Butterworth}},
  \bibinfo{author}{\bibfnamefont{A.~R.} \bibnamefont{Davison}},
  \bibinfo{author}{\bibfnamefont{M.}~\bibnamefont{Rubin}}, \bibnamefont{and}
  \bibinfo{author}{\bibfnamefont{G.~P.} \bibnamefont{Salam}},
  \bibinfo{journal}{Phys.Rev.Lett.} \textbf{\bibinfo{volume}{100}},
  \bibinfo{pages}{242001} (\bibinfo{year}{2008}{\natexlab{a}}),
  \eprint{0802.2470}.

\bibitem[{\citenamefont{Thaler and Wang}(2008)}]{Thaler:2008ju}
\bibinfo{author}{\bibfnamefont{J.}~\bibnamefont{Thaler}} \bibnamefont{and}
  \bibinfo{author}{\bibfnamefont{L.-T.} \bibnamefont{Wang}},
  \bibinfo{journal}{JHEP} \textbf{\bibinfo{volume}{0807}}, \bibinfo{pages}{092}
  (\bibinfo{year}{2008}), \eprint{0806.0023}.

\bibitem[{\citenamefont{Kaplan et~al.}(2008)\citenamefont{Kaplan, Rehermann,
  Schwartz, and Tweedie}}]{Kaplan:2008ie}
\bibinfo{author}{\bibfnamefont{D.~E.} \bibnamefont{Kaplan}},
  \bibinfo{author}{\bibfnamefont{K.}~\bibnamefont{Rehermann}},
  \bibinfo{author}{\bibfnamefont{M.~D.} \bibnamefont{Schwartz}},
  \bibnamefont{and} \bibinfo{author}{\bibfnamefont{B.}~\bibnamefont{Tweedie}},
  \bibinfo{journal}{Phys.Rev.Lett.} \textbf{\bibinfo{volume}{101}},
  \bibinfo{pages}{142001} (\bibinfo{year}{2008}), \eprint{0806.0848}.

\bibitem[{\citenamefont{Butterworth et~al.}(2009)\citenamefont{Butterworth,
  Davison, Rubin, and Salam}}]{Butterworth:2008sd}
\bibinfo{author}{\bibfnamefont{J.~M.} \bibnamefont{Butterworth}},
  \bibinfo{author}{\bibfnamefont{A.~R.} \bibnamefont{Davison}},
  \bibinfo{author}{\bibfnamefont{M.}~\bibnamefont{Rubin}}, \bibnamefont{and}
  \bibinfo{author}{\bibfnamefont{G.~P.} \bibnamefont{Salam}},
  \bibinfo{journal}{AIP Conf.Proc.} \textbf{\bibinfo{volume}{1078}},
  \bibinfo{pages}{189} (\bibinfo{year}{2009}), \eprint{0809.2530}.

\bibitem[{\citenamefont{Butterworth
  et~al.}(2008{\natexlab{b}})\citenamefont{Butterworth, Davison, Rubin, and
  Salam}}]{Butterworth:2008tr}
\bibinfo{author}{\bibfnamefont{J.~M.} \bibnamefont{Butterworth}},
  \bibinfo{author}{\bibfnamefont{A.~R.} \bibnamefont{Davison}},
  \bibinfo{author}{\bibfnamefont{M.}~\bibnamefont{Rubin}}, \bibnamefont{and}
  \bibinfo{author}{\bibfnamefont{G.~P.} \bibnamefont{Salam}}
  (\bibinfo{year}{2008}{\natexlab{b}}), \eprint{0810.0409}.

\bibitem[{\citenamefont{Ellis et~al.}(2009)\citenamefont{Ellis, Vermilion, and
  Walsh}}]{Ellis:2009su}
\bibinfo{author}{\bibfnamefont{S.~D.} \bibnamefont{Ellis}},
  \bibinfo{author}{\bibfnamefont{C.~K.} \bibnamefont{Vermilion}},
  \bibnamefont{and} \bibinfo{author}{\bibfnamefont{J.~R.} \bibnamefont{Walsh}},
  \bibinfo{journal}{Phys.Rev.} \textbf{\bibinfo{volume}{D80}},
  \bibinfo{pages}{051501} (\bibinfo{year}{2009}), \eprint{0903.5081}.

\bibitem[{\citenamefont{Ellis et~al.}(2010)\citenamefont{Ellis, Vermilion, and
  Walsh}}]{Ellis:2009me}
\bibinfo{author}{\bibfnamefont{S.~D.} \bibnamefont{Ellis}},
  \bibinfo{author}{\bibfnamefont{C.~K.} \bibnamefont{Vermilion}},
  \bibnamefont{and} \bibinfo{author}{\bibfnamefont{J.~R.} \bibnamefont{Walsh}},
  \bibinfo{journal}{Phys.Rev.} \textbf{\bibinfo{volume}{D81}},
  \bibinfo{pages}{094023} (\bibinfo{year}{2010}), \eprint{0912.0033}.

\bibitem[{\citenamefont{Krohn et~al.}(2010)\citenamefont{Krohn, Thaler, and
  Wang}}]{Krohn:2009th}
\bibinfo{author}{\bibfnamefont{D.}~\bibnamefont{Krohn}},
  \bibinfo{author}{\bibfnamefont{J.}~\bibnamefont{Thaler}}, \bibnamefont{and}
  \bibinfo{author}{\bibfnamefont{L.-T.} \bibnamefont{Wang}},
  \bibinfo{journal}{JHEP} \textbf{\bibinfo{volume}{1002}}, \bibinfo{pages}{084}
  (\bibinfo{year}{2010}), \eprint{0912.1342}.

\bibitem[{ATL(2012{\natexlab{b}})}]{ATLAS-CONF-2012-065}
\bibinfo{type}{Tech. Rep.} \bibinfo{number}{ATLAS-CONF-2012-065},
  \bibinfo{institution}{CERN}, \bibinfo{address}{Geneva}
  (\bibinfo{year}{2012}{\natexlab{b}}).

\bibitem[{CMS(2012)}]{CMS-PAS-EXO-11-095}
\bibinfo{type}{Tech. Rep.} \bibinfo{number}{CMS-PAS-EXO-11-095},
  \bibinfo{institution}{CERN}, \bibinfo{address}{Geneva}
  (\bibinfo{year}{2012}).

\bibitem[{\citenamefont{Plehn et~al.}(2010)\citenamefont{Plehn, Salam, and
  Spannowsky}}]{Plehn:2009rk}
\bibinfo{author}{\bibfnamefont{T.}~\bibnamefont{Plehn}},
  \bibinfo{author}{\bibfnamefont{G.~P.} \bibnamefont{Salam}}, \bibnamefont{and}
  \bibinfo{author}{\bibfnamefont{M.}~\bibnamefont{Spannowsky}},
  \bibinfo{journal}{Phys.Rev.Lett.} \textbf{\bibinfo{volume}{104}},
  \bibinfo{pages}{111801} (\bibinfo{year}{2010}), \eprint{0910.5472}.

\bibitem[{\citenamefont{Gallicchio et~al.}(2011)\citenamefont{Gallicchio, Huth,
  Kagan, Schwartz, Black et~al.}}]{Gallicchio:2010dq}
\bibinfo{author}{\bibfnamefont{J.}~\bibnamefont{Gallicchio}},
  \bibinfo{author}{\bibfnamefont{J.}~\bibnamefont{Huth}},
  \bibinfo{author}{\bibfnamefont{M.}~\bibnamefont{Kagan}},
  \bibinfo{author}{\bibfnamefont{M.~D.} \bibnamefont{Schwartz}},
  \bibinfo{author}{\bibfnamefont{K.}~\bibnamefont{Black}},
  \bibnamefont{et~al.}, \bibinfo{journal}{JHEP}
  \textbf{\bibinfo{volume}{1104}}, \bibinfo{pages}{069} (\bibinfo{year}{2011}),
  \eprint{1010.3698}.

\bibitem[{\citenamefont{Hackstein and Spannowsky}(2010)}]{Hackstein:2010wk}
\bibinfo{author}{\bibfnamefont{C.}~\bibnamefont{Hackstein}} \bibnamefont{and}
  \bibinfo{author}{\bibfnamefont{M.}~\bibnamefont{Spannowsky}},
  \bibinfo{journal}{Phys.Rev.} \textbf{\bibinfo{volume}{D82}},
  \bibinfo{pages}{113012} (\bibinfo{year}{2010}), \eprint{1008.2202}.

\bibitem[{\citenamefont{Kribs et~al.}(2010{\natexlab{a}})\citenamefont{Kribs,
  Martin, Roy, and Spannowsky}}]{Kribs:2009yh}
\bibinfo{author}{\bibfnamefont{G.~D.} \bibnamefont{Kribs}},
  \bibinfo{author}{\bibfnamefont{A.}~\bibnamefont{Martin}},
  \bibinfo{author}{\bibfnamefont{T.~S.} \bibnamefont{Roy}}, \bibnamefont{and}
  \bibinfo{author}{\bibfnamefont{M.}~\bibnamefont{Spannowsky}},
  \bibinfo{journal}{Phys.Rev.} \textbf{\bibinfo{volume}{D81}},
  \bibinfo{pages}{111501} (\bibinfo{year}{2010}{\natexlab{a}}),
  \eprint{0912.4731}.

\bibitem[{\citenamefont{Kribs et~al.}(2010{\natexlab{b}})\citenamefont{Kribs,
  Martin, Roy, and Spannowsky}}]{Kribs:2010hp}
\bibinfo{author}{\bibfnamefont{G.~D.} \bibnamefont{Kribs}},
  \bibinfo{author}{\bibfnamefont{A.}~\bibnamefont{Martin}},
  \bibinfo{author}{\bibfnamefont{T.~S.} \bibnamefont{Roy}}, \bibnamefont{and}
  \bibinfo{author}{\bibfnamefont{M.}~\bibnamefont{Spannowsky}},
  \bibinfo{journal}{Phys.Rev.} \textbf{\bibinfo{volume}{D82}},
  \bibinfo{pages}{095012} (\bibinfo{year}{2010}{\natexlab{b}}),
  \eprint{1006.1656}.

\bibitem[{\citenamefont{Kribs et~al.}(2011)\citenamefont{Kribs, Martin, and
  Roy}}]{Kribs:2010ii}
\bibinfo{author}{\bibfnamefont{G.~D.} \bibnamefont{Kribs}},
  \bibinfo{author}{\bibfnamefont{A.}~\bibnamefont{Martin}}, \bibnamefont{and}
  \bibinfo{author}{\bibfnamefont{T.~S.} \bibnamefont{Roy}},
  \bibinfo{journal}{Phys.Rev.} \textbf{\bibinfo{volume}{D84}},
  \bibinfo{pages}{095024} (\bibinfo{year}{2011}), \eprint{1012.2866}.

\bibitem[{\citenamefont{Katz et~al.}(2011)\citenamefont{Katz, Son, and
  Tweedie}}]{Katz:2010iq}
\bibinfo{author}{\bibfnamefont{A.}~\bibnamefont{Katz}},
  \bibinfo{author}{\bibfnamefont{M.}~\bibnamefont{Son}}, \bibnamefont{and}
  \bibinfo{author}{\bibfnamefont{B.}~\bibnamefont{Tweedie}},
  \bibinfo{journal}{Phys.Rev.} \textbf{\bibinfo{volume}{D83}},
  \bibinfo{pages}{114033} (\bibinfo{year}{2011}), \eprint{1011.4523}.

\bibitem[{\citenamefont{Englert et~al.}(2011)\citenamefont{Englert, Roy, and
  Spannowsky}}]{Englert:2011iz}
\bibinfo{author}{\bibfnamefont{C.}~\bibnamefont{Englert}},
  \bibinfo{author}{\bibfnamefont{T.~S.} \bibnamefont{Roy}}, \bibnamefont{and}
  \bibinfo{author}{\bibfnamefont{M.}~\bibnamefont{Spannowsky}},
  \bibinfo{journal}{Phys.Rev.} \textbf{\bibinfo{volume}{D84}},
  \bibinfo{pages}{075026} (\bibinfo{year}{2011}), \eprint{1106.4545}.

\bibitem[{\citenamefont{Son et~al.}(2012)\citenamefont{Son, Spethmann, and
  Tweedie}}]{Son:2012mb}
\bibinfo{author}{\bibfnamefont{M.}~\bibnamefont{Son}},
  \bibinfo{author}{\bibfnamefont{C.}~\bibnamefont{Spethmann}},
  \bibnamefont{and} \bibinfo{author}{\bibfnamefont{B.}~\bibnamefont{Tweedie}}
  (\bibinfo{year}{2012}), \eprint{1204.0525}.

\bibitem[{\citenamefont{Abdesselam et~al.}(2011)\citenamefont{Abdesselam,
  Kuutmann, Bitenc, Brooijmans, Butterworth et~al.}}]{Abdesselam:2010pt}
\bibinfo{author}{\bibfnamefont{A.}~\bibnamefont{Abdesselam}},
  \bibinfo{author}{\bibfnamefont{E.~B.} \bibnamefont{Kuutmann}},
  \bibinfo{author}{\bibfnamefont{U.}~\bibnamefont{Bitenc}},
  \bibinfo{author}{\bibfnamefont{G.}~\bibnamefont{Brooijmans}},
  \bibinfo{author}{\bibfnamefont{J.}~\bibnamefont{Butterworth}},
  \bibnamefont{et~al.}, \bibinfo{journal}{Eur.Phys.J.}
  \textbf{\bibinfo{volume}{C71}}, \bibinfo{pages}{1661} (\bibinfo{year}{2011}),
  \eprint{1012.5412}.

\bibitem[{\citenamefont{Altheimer et~al.}(2012)\citenamefont{Altheimer, Arora,
  Asquith, Brooijmans, Butterworth et~al.}}]{Altheimer:2012mn}
\bibinfo{author}{\bibfnamefont{A.}~\bibnamefont{Altheimer}},
  \bibinfo{author}{\bibfnamefont{S.}~\bibnamefont{Arora}},
  \bibinfo{author}{\bibfnamefont{L.}~\bibnamefont{Asquith}},
  \bibinfo{author}{\bibfnamefont{G.}~\bibnamefont{Brooijmans}},
  \bibinfo{author}{\bibfnamefont{J.}~\bibnamefont{Butterworth}},
  \bibnamefont{et~al.}, \bibinfo{journal}{J.Phys.G}
  \textbf{\bibinfo{volume}{G39}}, \bibinfo{pages}{063001}
  (\bibinfo{year}{2012}), \eprint{1201.0008}.

\bibitem[{\citenamefont{Alves et~al.}(2011)}]{Alves:2011wf}
\bibinfo{author}{\bibfnamefont{D.}~\bibnamefont{Alves}} \bibnamefont{et~al.}
  (\bibinfo{collaboration}{LHC New Physics Working Group})
  (\bibinfo{year}{2011}), \eprint{1105.2838}.

\bibitem[{\citenamefont{Schabinger and Wells}(2005)}]{Schabinger:2005ei}
\bibinfo{author}{\bibfnamefont{R.}~\bibnamefont{Schabinger}} \bibnamefont{and}
  \bibinfo{author}{\bibfnamefont{J.~D.} \bibnamefont{Wells}},
  \bibinfo{journal}{Phys.Rev.} \textbf{\bibinfo{volume}{D72}},
  \bibinfo{pages}{093007} (\bibinfo{year}{2005}), \eprint{hep-ph/0509209}.

\bibitem[{\citenamefont{Patt and Wilczek}(2006)}]{Patt:2006fw}
\bibinfo{author}{\bibfnamefont{B.}~\bibnamefont{Patt}} \bibnamefont{and}
  \bibinfo{author}{\bibfnamefont{F.}~\bibnamefont{Wilczek}}
  (\bibinfo{year}{2006}), \eprint{hep-ph/0605188}.

\bibitem[{\citenamefont{Strassler and Zurek}(2007)}]{Strassler:2006im}
\bibinfo{author}{\bibfnamefont{M.~J.} \bibnamefont{Strassler}}
  \bibnamefont{and} \bibinfo{author}{\bibfnamefont{K.~M.} \bibnamefont{Zurek}},
  \bibinfo{journal}{Phys.Lett.} \textbf{\bibinfo{volume}{B651}},
  \bibinfo{pages}{374} (\bibinfo{year}{2007}), \eprint{hep-ph/0604261}.

\bibitem[{\citenamefont{Alwall et~al.}(2011)\citenamefont{Alwall, Herquet,
  Maltoni, Mattelaer, and Stelzer}}]{Alwall:2011uj}
\bibinfo{author}{\bibfnamefont{J.}~\bibnamefont{Alwall}},
  \bibinfo{author}{\bibfnamefont{M.}~\bibnamefont{Herquet}},
  \bibinfo{author}{\bibfnamefont{F.}~\bibnamefont{Maltoni}},
  \bibinfo{author}{\bibfnamefont{O.}~\bibnamefont{Mattelaer}},
  \bibnamefont{and} \bibinfo{author}{\bibfnamefont{T.}~\bibnamefont{Stelzer}},
  \bibinfo{journal}{JHEP} \textbf{\bibinfo{volume}{1106}}, \bibinfo{pages}{128}
  (\bibinfo{year}{2011}), \eprint{1106.0522}.

\bibitem[{\citenamefont{Sjostrand et~al.}(2006)\citenamefont{Sjostrand, Mrenna,
  and Skands}}]{Sjostrand:2006za}
\bibinfo{author}{\bibfnamefont{T.}~\bibnamefont{Sjostrand}},
  \bibinfo{author}{\bibfnamefont{S.}~\bibnamefont{Mrenna}}, \bibnamefont{and}
  \bibinfo{author}{\bibfnamefont{P.~Z.} \bibnamefont{Skands}},
  \bibinfo{journal}{JHEP} \textbf{\bibinfo{volume}{0605}}, \bibinfo{pages}{026}
  (\bibinfo{year}{2006}), \eprint{hep-ph/0603175}.

\bibitem[{\citenamefont{Sjostrand et~al.}(2008)\citenamefont{Sjostrand, Mrenna,
  and Skands}}]{Sjostrand:2007gs}
\bibinfo{author}{\bibfnamefont{T.}~\bibnamefont{Sjostrand}},
  \bibinfo{author}{\bibfnamefont{S.}~\bibnamefont{Mrenna}}, \bibnamefont{and}
  \bibinfo{author}{\bibfnamefont{P.~Z.} \bibnamefont{Skands}},
  \bibinfo{journal}{Comput.Phys.Commun.} \textbf{\bibinfo{volume}{178}},
  \bibinfo{pages}{852} (\bibinfo{year}{2008}), \eprint{0710.3820}.

\bibitem[{\citenamefont{private correspondence~with
  Henry~Lubatti.}(2012)}]{Smearing}
\bibinfo{author}{\bibnamefont{private correspondence~with Henry~Lubatti.}}


\bibitem[{\citenamefont{Cacciari and Salam}(2006)}]{Cacciari:2005hq}
\bibinfo{author}{\bibfnamefont{M.}~\bibnamefont{Cacciari}} \bibnamefont{and}
  \bibinfo{author}{\bibfnamefont{G.~P.} \bibnamefont{Salam}},
  \bibinfo{journal}{Phys.Lett.} \textbf{\bibinfo{volume}{B641}},
  \bibinfo{pages}{57} (\bibinfo{year}{2006}), \eprint{hep-ph/0512210}.

\bibitem[{\citenamefont{Cacciari et~al.}(2012)\citenamefont{Cacciari, Salam,
  and Soyez}}]{Cacciari:2011ma}
\bibinfo{author}{\bibfnamefont{M.}~\bibnamefont{Cacciari}},
  \bibinfo{author}{\bibfnamefont{G.~P.} \bibnamefont{Salam}}, \bibnamefont{and}
  \bibinfo{author}{\bibfnamefont{G.}~\bibnamefont{Soyez}},
  \bibinfo{journal}{Eur.Phys.J.} \textbf{\bibinfo{volume}{C72}},
  \bibinfo{pages}{1896} (\bibinfo{year}{2012}), \eprint{1111.6097}.

\bibitem[{\citenamefont{Cacciari
  et~al.}(2008{\natexlab{a}})\citenamefont{Cacciari, Salam, and
  Soyez}}]{Cacciari:2008gp}
\bibinfo{author}{\bibfnamefont{M.}~\bibnamefont{Cacciari}},
  \bibinfo{author}{\bibfnamefont{G.~P.} \bibnamefont{Salam}}, \bibnamefont{and}
  \bibinfo{author}{\bibfnamefont{G.}~\bibnamefont{Soyez}},
  \bibinfo{journal}{JHEP} \textbf{\bibinfo{volume}{0804}}, \bibinfo{pages}{063}
  (\bibinfo{year}{2008}{\natexlab{a}}), \eprint{0802.1189}.

\bibitem[{\citenamefont{Catani et~al.}(1993)\citenamefont{Catani, Dokshitzer,
  Seymour, and Webber}}]{Catani:1993hr}
\bibinfo{author}{\bibfnamefont{S.}~\bibnamefont{Catani}},
  \bibinfo{author}{\bibfnamefont{Y.~L.} \bibnamefont{Dokshitzer}},
  \bibinfo{author}{\bibfnamefont{M.}~\bibnamefont{Seymour}}, \bibnamefont{and}
  \bibinfo{author}{\bibfnamefont{B.}~\bibnamefont{Webber}},
  \bibinfo{journal}{Nucl.Phys.} \textbf{\bibinfo{volume}{B406}},
  \bibinfo{pages}{187} (\bibinfo{year}{1993}).

\bibitem[{\citenamefont{Ellis and Soper}(1993)}]{Ellis:1993tq}
\bibinfo{author}{\bibfnamefont{S.~D.} \bibnamefont{Ellis}} \bibnamefont{and}
  \bibinfo{author}{\bibfnamefont{D.~E.} \bibnamefont{Soper}},
  \bibinfo{journal}{Phys.Rev.} \textbf{\bibinfo{volume}{D48}},
  \bibinfo{pages}{3160} (\bibinfo{year}{1993}), \eprint{hep-ph/9305266}.

\bibitem[{\citenamefont{Dokshitzer et~al.}(1997)\citenamefont{Dokshitzer,
  Leder, Moretti, and Webber}}]{Dokshitzer:1997in}
\bibinfo{author}{\bibfnamefont{Y.~L.} \bibnamefont{Dokshitzer}},
  \bibinfo{author}{\bibfnamefont{G.}~\bibnamefont{Leder}},
  \bibinfo{author}{\bibfnamefont{S.}~\bibnamefont{Moretti}}, \bibnamefont{and}
  \bibinfo{author}{\bibfnamefont{B.}~\bibnamefont{Webber}},
  \bibinfo{journal}{JHEP} \textbf{\bibinfo{volume}{9708}}, \bibinfo{pages}{001}
  (\bibinfo{year}{1997}), \eprint{hep-ph/9707323}.

\bibitem[{\citenamefont{Wobisch and Wengler}(1998)}]{Wobisch:1998wt}
\bibinfo{author}{\bibfnamefont{M.}~\bibnamefont{Wobisch}} \bibnamefont{and}
  \bibinfo{author}{\bibfnamefont{T.}~\bibnamefont{Wengler}}
  (\bibinfo{year}{1998}), \eprint{hep-ph/9907280}.

\bibitem[{\citenamefont{Wobisch}(2000)}]{Wobisch:2000dk}
\bibinfo{author}{\bibfnamefont{M.}~\bibnamefont{Wobisch}}
  (\bibinfo{year}{2000}).

\bibitem[{\citenamefont{Thaler and Van~Tilburg}(2011)}]{Thaler:2010tr}
\bibinfo{author}{\bibfnamefont{J.}~\bibnamefont{Thaler}} \bibnamefont{and}
  \bibinfo{author}{\bibfnamefont{K.}~\bibnamefont{Van~Tilburg}},
  \bibinfo{journal}{JHEP} \textbf{\bibinfo{volume}{1103}}, \bibinfo{pages}{015}
  (\bibinfo{year}{2011}), \eprint{1011.2268}.

\bibitem[{\citenamefont{Thaler and Van~Tilburg}(2012)}]{Thaler:2011gf}
\bibinfo{author}{\bibfnamefont{J.}~\bibnamefont{Thaler}} \bibnamefont{and}
  \bibinfo{author}{\bibfnamefont{K.}~\bibnamefont{Van~Tilburg}},
  \bibinfo{journal}{JHEP} \textbf{\bibinfo{volume}{1202}}, \bibinfo{pages}{093}
  (\bibinfo{year}{2012}), \eprint{1108.2701}.

\bibitem[{\citenamefont{Stewart et~al.}(2010)\citenamefont{Stewart, Tackmann,
  and Waalewijn}}]{Stewart:2010tn}
\bibinfo{author}{\bibfnamefont{I.~W.} \bibnamefont{Stewart}},
  \bibinfo{author}{\bibfnamefont{F.~J.} \bibnamefont{Tackmann}},
  \bibnamefont{and} \bibinfo{author}{\bibfnamefont{W.~J.}
  \bibnamefont{Waalewijn}}, \bibinfo{journal}{Phys.Rev.Lett.}
  \textbf{\bibinfo{volume}{105}}, \bibinfo{pages}{092002}
  (\bibinfo{year}{2010}), \eprint{1004.2489}.

\bibitem[{\citenamefont{Ellis et~al.}(2012{\natexlab{a}})\citenamefont{Ellis,
  Hornig, Roy, Krohn, and Schwartz}}]{Ellis:2012sn}
\bibinfo{author}{\bibfnamefont{S.~D.} \bibnamefont{Ellis}},
  \bibinfo{author}{\bibfnamefont{A.}~\bibnamefont{Hornig}},
  \bibinfo{author}{\bibfnamefont{T.~S.} \bibnamefont{Roy}},
  \bibinfo{author}{\bibfnamefont{D.}~\bibnamefont{Krohn}}, \bibnamefont{and}
  \bibinfo{author}{\bibfnamefont{M.~D.} \bibnamefont{Schwartz}},
  \bibinfo{journal}{Phys.Rev.Lett.} \textbf{\bibinfo{volume}{108}},
  \bibinfo{pages}{182003} (\bibinfo{year}{2012}{\natexlab{a}}),
  \eprint{1201.1914},
  \urlprefix\url{http://jets.physics.harvard.edu/Qjets/html/Welcome.html}.

\bibitem[{\citenamefont{Cacciari
  et~al.}(2008{\natexlab{b}})\citenamefont{Cacciari, Salam, and
  Soyez}}]{Cacciari:2008gn}
\bibinfo{author}{\bibfnamefont{M.}~\bibnamefont{Cacciari}},
  \bibinfo{author}{\bibfnamefont{G.~P.} \bibnamefont{Salam}}, \bibnamefont{and}
  \bibinfo{author}{\bibfnamefont{G.}~\bibnamefont{Soyez}},
  \bibinfo{journal}{JHEP} \textbf{\bibinfo{volume}{0804}}, \bibinfo{pages}{005}
  (\bibinfo{year}{2008}{\natexlab{b}}), \eprint{0802.1188}.

\bibitem[{\citenamefont{Freund and Schapire}(1996)}]{BDT}
\bibinfo{author}{\bibfnamefont{Y.}~\bibnamefont{Freund}} \bibnamefont{and}
  \bibinfo{author}{\bibfnamefont{R.~E.} \bibnamefont{Schapire}}, pp.
  \bibinfo{pages}{148--156} (\bibinfo{year}{1996}).

\bibitem[{\citenamefont{Hocker et~al.}(2007)\citenamefont{Hocker, Stelzer,
  Tegenfeldt, Voss, Voss et~al.}}]{Hocker:2007ht}
\bibinfo{author}{\bibfnamefont{A.}~\bibnamefont{Hocker}},
  \bibinfo{author}{\bibfnamefont{J.}~\bibnamefont{Stelzer}},
  \bibinfo{author}{\bibfnamefont{F.}~\bibnamefont{Tegenfeldt}},
  \bibinfo{author}{\bibfnamefont{H.}~\bibnamefont{Voss}},
  \bibinfo{author}{\bibfnamefont{K.}~\bibnamefont{Voss}}, \bibnamefont{et~al.},
  \bibinfo{journal}{PoS} \textbf{\bibinfo{volume}{ACAT}}, \bibinfo{pages}{040}
  (\bibinfo{year}{2007}), \eprint{physics/0703039}.

\bibitem[{\citenamefont{Ellis et~al.}(2012{\natexlab{a}})\citenamefont{Ellis,
  Roy, and Scholtz}}]{Ellis:2012sd}
\bibinfo{author}{\bibfnamefont{S.~D.} \bibnamefont{Ellis}},
  \bibinfo{author}{\bibfnamefont{T.~S.} \bibnamefont{Roy}}, \bibnamefont{and}
  \bibinfo{author}{\bibfnamefont{J.}~\bibnamefont{Scholtz}}
  (\bibinfo{year}{2012}{\natexlab{a}}), \eprint{1210.1855}.

\bibitem[{\citenamefont{Ruderman and Volansky}(2010)}]{Ruderman:2009tj}
\bibinfo{author}{\bibfnamefont{J.~T.} \bibnamefont{Ruderman}} \bibnamefont{and}
  \bibinfo{author}{\bibfnamefont{T.}~\bibnamefont{Volansky}},
  \bibinfo{journal}{JHEP} \textbf{\bibinfo{volume}{1002}}, \bibinfo{pages}{024}
  (\bibinfo{year}{2010}), \eprint{0908.1570}.

\bibitem[{\citenamefont{Cheung et~al.}(2010)\citenamefont{Cheung, Ruderman,
  Wang, and Yavin}}]{Cheung:2009su}
\bibinfo{author}{\bibfnamefont{C.}~\bibnamefont{Cheung}},
  \bibinfo{author}{\bibfnamefont{J.~T.} \bibnamefont{Ruderman}},
  \bibinfo{author}{\bibfnamefont{L.-T.} \bibnamefont{Wang}}, \bibnamefont{and}
  \bibinfo{author}{\bibfnamefont{I.}~\bibnamefont{Yavin}},
  \bibinfo{journal}{JHEP} \textbf{\bibinfo{volume}{1004}}, \bibinfo{pages}{116}
  (\bibinfo{year}{2010}), \eprint{0909.0290}.

\bibitem[{\citenamefont{Falkowski
  et~al.}(2010{\natexlab{a}})\citenamefont{Falkowski, Ruderman, Volansky, and
  Zupan}}]{Falkowski:2010cm}
\bibinfo{author}{\bibfnamefont{A.}~\bibnamefont{Falkowski}},
  \bibinfo{author}{\bibfnamefont{J.~T.} \bibnamefont{Ruderman}},
  \bibinfo{author}{\bibfnamefont{T.}~\bibnamefont{Volansky}}, \bibnamefont{and}
  \bibinfo{author}{\bibfnamefont{J.}~\bibnamefont{Zupan}},
  \bibinfo{journal}{JHEP} \textbf{\bibinfo{volume}{1005}}, \bibinfo{pages}{077}
  (\bibinfo{year}{2010}{\natexlab{a}}), \eprint{1002.2952}.

\bibitem[{\citenamefont{Falkowski
  et~al.}(2010{\natexlab{b}})\citenamefont{Falkowski, Ruderman, Volansky, and
  Zupan}}]{Falkowski:2010gv}
\bibinfo{author}{\bibfnamefont{A.}~\bibnamefont{Falkowski}},
  \bibinfo{author}{\bibfnamefont{J.~T.} \bibnamefont{Ruderman}},
  \bibinfo{author}{\bibfnamefont{T.}~\bibnamefont{Volansky}}, \bibnamefont{and}
  \bibinfo{author}{\bibfnamefont{J.}~\bibnamefont{Zupan}},
  \bibinfo{journal}{Phys.Rev.Lett.} \textbf{\bibinfo{volume}{105}},
  \bibinfo{pages}{241801} (\bibinfo{year}{2010}{\natexlab{b}}),
  \eprint{1007.3496}.

\bibitem[{\citenamefont{Aad et~al.}(2008)}]{Aad:2008zzm}
\bibinfo{author}{\bibfnamefont{G.}~\bibnamefont{Aad}} \bibnamefont{et~al.}
  (\bibinfo{collaboration}{ATLAS Collaboration}), \bibinfo{journal}{JINST}
  \textbf{\bibinfo{volume}{3}}, \bibinfo{pages}{S08003} (\bibinfo{year}{2008}).

\bibitem[{\citenamefont{Conway et~al.}(2012)\citenamefont{Conway, Culbertson,
  Demina, Kilminster, Kruse, Mrenna, Nielsen, Roco, Pierce, Thaler
  et~al.}}]{PGS4}
\bibinfo{author}{\bibfnamefont{J.}~\bibnamefont{Conway}},
  \bibinfo{author}{\bibfnamefont{R.}~\bibnamefont{Culbertson}},
  \bibinfo{author}{\bibfnamefont{R.}~\bibnamefont{Demina}},
  \bibinfo{author}{\bibfnamefont{B.}~\bibnamefont{Kilminster}},
  \bibinfo{author}{\bibfnamefont{M.}~\bibnamefont{Kruse}},
  \bibinfo{author}{\bibfnamefont{S.}~\bibnamefont{Mrenna}},
  \bibinfo{author}{\bibfnamefont{J.}~\bibnamefont{Nielsen}},
  \bibinfo{author}{\bibfnamefont{M.}~\bibnamefont{Roco}},
  \bibinfo{author}{\bibfnamefont{A.}~\bibnamefont{Pierce}},
  \bibinfo{author}{\bibfnamefont{J.}~\bibnamefont{Thaler}},
  \bibnamefont{et~al.} (\bibinfo{year}{2012}),
  \urlprefix\url{http://www.physics.ucdavis.edu/~conway/research/software/pgs/pgs4-general.htm}.

\end{thebibliography}

\end{document}